\documentclass[12pt,nofootinbib]{article}
\usepackage[dvips]{graphicx}
\usepackage{amssymb}
\usepackage{amsmath}
\usepackage{epsfig}
\usepackage{graphicx}

\def\beq{\begin{equation}}
\def\eeq{\end{equation}}
\def\be{\begin{equation}}
\def\ee{\end{equation}}
\def\bea{\begin{eqnarray}}
\def\eea{\end{eqnarray}}
\def\d{{\rm d}}

\DeclareRobustCommand{\SkipTocEntry}[4]{}

\begin{document}


\begin{titlepage}

\setcounter{page}{1} \baselineskip=15.5pt \thispagestyle{empty}

\begin{flushright}
hep-th/0607050\\
PUPT-2202\\
ITEP-TH-32/06
\end{flushright}
\vfil

\begin{center}
{\LARGE On D3-brane Potentials in Compactifications} \vskip0.3cm
{\LARGE with Fluxes and Wrapped D-branes}

\end{center}
\bigskip\

\begin{center}
{\large Daniel Baumann,$^{1}$ Anatoly Dymarsky,$^{1}$ Igor R.
Klebanov,$^{1}$}
\vskip 3pt
{\large Juan Maldacena,$^{2}$ Liam McAllister,$^{1}$ and Arvind
Murugan$^{1}$}
\end{center}

\begin{center}
${}^1$\textit{Joseph Henry Laboratories, Princeton University,
Princeton, NJ 08544}
\vskip 3pt
${}^2$\textit{Institute for Advanced Study, Princeton, NJ 08540}
\end{center} \vfil


\noindent We study the potential governing D3-brane motion in a
warped throat region of a string compactification with internal
fluxes and wrapped D-branes. If the K\"ahler moduli of the compact
space are stabilized by nonperturbative effects, a D3-brane
experiences a force due to its interaction with D-branes wrapping
certain four-cycles. We compute this interaction, as a correction
to the warped four-cycle volume, using explicit throat backgrounds
in supergravity. This amounts to a closed-string channel
computation of the loop corrections to the nonperturbative
superpotential that stabilizes the volume. We demonstrate for
warped conical spaces that the superpotential correction is given
by the embedding equation specifying the wrapped four-cycle, in
agreement with the general form proposed by Ganor.  Our approach
automatically provides a solution to the problem of defining a
holomorphic gauge coupling on wrapped D7-branes in a background
with D3-branes. Finally, our results have applications to
cosmological inflation models in which the inflaton is modeled by
a D3-brane moving in a warped throat.

\vfil
\begin{flushleft}
\today
\end{flushleft}

\end{titlepage}

\newpage


\tableofcontents
\newpage


\section{Introduction}
\label{sec:intro}

\addtocontents{toc}{\SkipTocEntry}
\subsection{Motivation}

Cosmological inflation \cite{Inflation} is a remarkable idea that
provides a convincing explanation for the isotropy and homogeneity
of the universe. In addition, the theory contains an elegant
mechanism to account for the quantum origin of large scale
structure. The observational evidence for inflation is strong and
rapidly growing \cite{Observations}, and in the near future it
will be possible to falsify a large fraction of existing models.
This presents a remarkable opportunity for inflationary
model-building, and it intensifies the need for a more fundamental
description of inflation than current phenomenological models can
provide.

In string theory, considerable effort has been devoted to this
problem. One promising idea is the identification of the inflaton
field with the internal coordinate of a mobile D3-brane, as in
brane-antibrane inflation models \cite{Dvali,otherbraneinflation},
in which a Coulombic interaction between the branes gives rise to
the inflaton potential. At the same time, advances in string
compactification \cite{GKP,KKLT} (for reviews, see \cite{reviews})
have enabled the construction of solutions in which all moduli are
stabilized by a combination of internal fluxes and wrapped
D-branes. This has led to the formulation of realistic and
moderately explicit models in which the brane-antibrane pair is
inserted into such a stabilized flux compactification
\cite{KKLMMT,otherstringinflation,Tye}. Particularly in warped
throat regions of the compact space, the force between the branes
can be weak enough to allow for prolonged inflation.  It is
therefore interesting to study the detailed potential determining
D3-brane motion in a warped throat region, such as the warped
deformed conifold \cite{KS} or its `baryonic branch'
generalizations, the resolved warped deformed conifolds
\cite{Butti,Dymarsky:2005xt}. In \cite{Dymarsky:2005xt} it was
observed that a mobile D3-brane in a resolved warped deformed
conifold experiences a force even in the absence of an antibrane
at the bottom of the throat. This makes \cite{Dymarsky:2005xt} a
possible alternative to the brane-antibrane scenario of
\cite{KKLMMT}. The calculations in this paper are carried out in a
region sufficiently far from the bottom of the throat that the
metrics of \cite{KS,Butti,Dymarsky:2005xt} are well-approximated
by the asymptotic warped conifold metric found in \cite{KT}.
Therefore, our methods apply to both scenarios, as well as to
their generalizations to other warped cones.

A truly satisfactory model of inflation in string theory should
include a complete specification of a string compactification,
together with a reliable computation of the resulting
four-dimensional effective theory.  While some models come close
to this goal, very small corrections to the potential can spoil
the delicate flatness conditions required for slow-roll inflation
\cite{SlowRoll}.  In particular, gravitational corrections
typically induce inflaton masses of order the Hubble parameter
$H$, which are fatal for slow-roll.  String theory provides a
framework for a systematic computation of these corrections, but
so far it has rarely been possible, in practice, to compute all
the relevant effects. However, there is no obstacle in principle,
and one of our main goals in this work is to improve the status of
this problem.

It is well-known that a D3-brane probe of a `no-scale'
compactification \cite{GKP} with imaginary self-dual three-form
fluxes experiences no force: gravitational attraction and
Ramond-Ramond repulsion cancel, and the brane can sit at any point
of the compact space with no energy cost. This no-force result is
no longer true, in general, when the volume of the
compactification is stabilized.  The D-brane moduli space is
lifted by the same nonperturbative effect that fixes the
compactification volume.  This has particular relevance for
inflation models involving moving D-branes.

In the warped brane inflation model of Kachru {\it{et al.}}
\cite{KKLMMT} it was established that the interaction potential of
a brane-antibrane pair in a warped throat geometry is
exceptionally flat, in the approximation that moduli-stabilization
effects are neglected. However, incorporating these effects
yielded a potential that generically was not flat enough for slow
roll. That is, certain correction terms to the inflaton potential
arising from the K\"ahler potential\footnote{These terms are those
associated with the usual supergravity eta problem.} and from
volume-inflaton mixing \cite{KKLMMT} could be computed in detail,
and gave explicit inflaton masses of order $H$.\footnote{ Similar
problems are expected to affect other warped throat inflation
scenarios, such as \cite{Dymarsky:2005xt}. Indeed, concerns about
the Hubble-scale corrections to the inflaton potential of
\cite{Dymarsky:2005xt} have been raised in \cite{Buchel:2006em},
but the effects of compactification were not considered there. In
this paper we calculate some of the most salient such effects,
those due to D-branes wrapping internal four-cycles. }  One
further mass term, arising from a one-loop correction to the
volume-stabilizing nonperturbative superpotential, was shown to be
present, but was not computed. The authors of \cite{KKLMMT} argued
that in some small percentage of possible models, this one-loop
mass term might take a value that approximately canceled the other
inflaton mass terms and produced an overall potential suitable for
slow-roll. This was a fine-tuning, but {\it{not}} an explicit one:
lacking a concrete computation of the one-loop correction, it was
not possible to specify fine-tuned microscopic parameters, such as
fluxes, geometry, and brane locations, in such a way that the
total mass term was known to be small. In this paper we give an
explicit computation of this key, missing inflaton mass term for
brane motion in general warped throat backgrounds. Applications of
our results to issues in brane inflation will be discussed in a
future paper \cite{PaperII}.

\addtocontents{toc}{\SkipTocEntry}
\subsection{Method}

The inflaton mass problem described in \cite{KKLMMT} appears in
any model of slow-roll inflation involving D3-branes moving in a
stabilized flux compactification. Thus, it is necessary to search
for a general method for computing the dependence of the
nonperturbative superpotential on the D3-brane position. Ganor
\cite{Ganor} studied this problem early on, and found that the
correction to the superpotential is a section of a bundle called
the `divisor bundle', which has a zero at the four-cycle where the
wrapped brane is located. The problem was addressed more
explicitly by Berg, Haack, and K\"ors (BHK) \cite{BHK}, who
computed the threshold corrections to gaugino condensate
superpotentials in toroidal orientifolds. This gave a
substantially complete\footnote{Corrections to the K\"ahler
potential provide one additional effect; see
\cite{Hebecker,BHK2}.} potential for brane inflation models in
such backgrounds. However, their approach involved a challenging
open-string one-loop computation that is difficult to generalize
to more complicated Calabi-Yau geometries and to backgrounds with
flux and warping, such as the warped throat backgrounds relevant
for a sizeable fraction of current models. Moreover, KKLT-type
volume stabilization often proceeds via a superpotential generated
by Euclidean D3-branes \cite{Witten}, not by gaugino condensation
or other strong gauge dynamics.  In this case the one-loop
correction comes from an instanton fluctuation determinant, which
has not been computed to date.

Following \cite{Giddings}, we overcome these difficulties by
viewing the correction to the mobile D3-brane potential as arising
from a distortion, sourced by the D3-brane itself, of the
background near the four-cycle responsible for the nonperturbative
effect.  This corrects the warped volume of the four-cycle,
changing the magnitude of the nonperturbative effect.
Specifically, we assume that the K\"ahler moduli are stabilized by
nonperturbative effects, arising either from Euclidean D3-branes
or from strong gauge dynamics (such as gaugino condensation) on
D7-branes.  In either case, the nonperturbative superpotential is
associated with a particular four-cycle, and has exponential
dependence on the warped volume of this cycle.  Inclusion of a
D3-brane in the compact space slightly modifies the supergravity
background, changing the warped volume of the four-cycle and hence
the gauge coupling in the D7-brane gauge theory. Due to gaugino
condensation this in turn changes the superpotential of the
four-dimensional effective theory. The result is an energy cost
for the D3-brane that depends on its location.

This method may be viewed as a closed-string dual of the
open-string computation of BHK \cite{BHK}.  In \S\ref{sec:BHK} we
compute the correction for a toroidal compactification, where an
explicit comparison is possible, and verify that the closed-string
method exactly reproduces the result of \cite{BHK}.  We view this
as a highly nontrivial check of the closed-string method.

Employing the closed-string perspective allows us to study the
potential for a D3-brane in a warped throat region, such as the
warped deformed conifold \cite{KS} or its generalizations
\cite{Butti,Dymarsky:2005xt}, glued into a flux compactification.
This is a case of direct phenomenological interest.  To model the
four-cycle bearing the most relevant nonperturbative effect, we
compute the change in the warped volume of a variety of
holomorphic four-cycles, as a function of the D3-brane position.
We find that most of the details of the geometry far from the
throat region are irrelevant.  Note that our method is applicable
provided that the internal manifold has large volume.

The distortion produced by moving a D3-brane in a warped throat
corresponds to a deformation of the gauge theory dual to
the throat by expectation values of certain
gauge-invariant operators \cite{KW2}. Hence, it is possible, and
convenient, to use methods and perspectives from the
AdS/CFT correspondence \cite{AdSCFT} (see \cite{MAGOO,IRK} for
reviews).

\addtocontents{toc}{\SkipTocEntry}
\subsection{Outline}

The organization of this paper is as follows.  In
\S\ref{sec:review} we recall the problem of determining the
potential for a D3-brane in a stabilized flux compactification. We
stress that a consistent computation must include a one-loop
correction to the volume-stabilizing nonperturbative
superpotential.  In \S\ref{sec:backreaction} we explain how this
correction may be computed in supergravity, as a correction to the
warped volume of each four-cycle producing a nonperturbative
effect. We present the Green's function method for determining the
perturbation of the warp factor at the location of the four-cycle
in \S\ref{sec:GFM}. We argue that supersymmetric four-cycles
provide a good model for the four-cycles producing nonperturbative
effects in general compactifications, and in particular in warped
throats.  In \S\ref{sec:conifold} we compute in detail the
corrected warped volumes of certain supersymmetric four-cycles in
the singular conifold. We also give results for corrected volumes
in some other asymptotically conical spaces.  In
\S\ref{sec:compactification} we give an explicit and physically
intuitive solution to the `rho problem' \cite{BHK}, {\it{i.e.}}
the problem of defining a holomorphic volume modulus in a
compactification with D3-branes. We also discuss the important
possibility of model-dependent effects from the bulk of the
compactification. We conclude in \S\ref{sec:conclusion}.

In Appendix \ref{sec:GF} we present some facts about Green's
functions on conical geometries, as needed for the computation of
\S\ref{sec:conifold}. We relegate the technical details of our
computation for warped conifolds to Appendix \ref{sec:Calc}. The
equivalent calculation for $Y^{p,q}$ cones is presented in
Appendix \ref{sec:ypq}.

\section{D3-branes and Volume Stabilization}
\label{sec:review}

\subsection{Nonperturbative Volume Stabilization}
\label{subsec:np}

For realistic applications to cosmology and particle
phenomenology, it is important to stabilize all the moduli.  The
flux-induced superpotential \cite{GVW} stabilizes the dilaton and
the complex structure moduli \cite{GKP}, but is independent of the
K\"ahler moduli. However, nonperturbative terms in the
superpotential do depend on the K\"ahler moduli, and hence can
lead to their stabilization \cite{KKLT}. There are two sources for
such effects:

\begin{enumerate}

\item{Euclidean D3-branes wrapping a four-cycle in the Calabi-Yau
\cite{Witten}.}

\item{Gaugino condensation or other strong gauge dynamics on a stack
of $N_{D7}$ spacetime-filling D7-branes wrapping a four-cycle in
the Calabi-Yau. }

\end{enumerate}

\begin{figure}[htbp]
    \centering
        \includegraphics[width=0.80\textwidth]{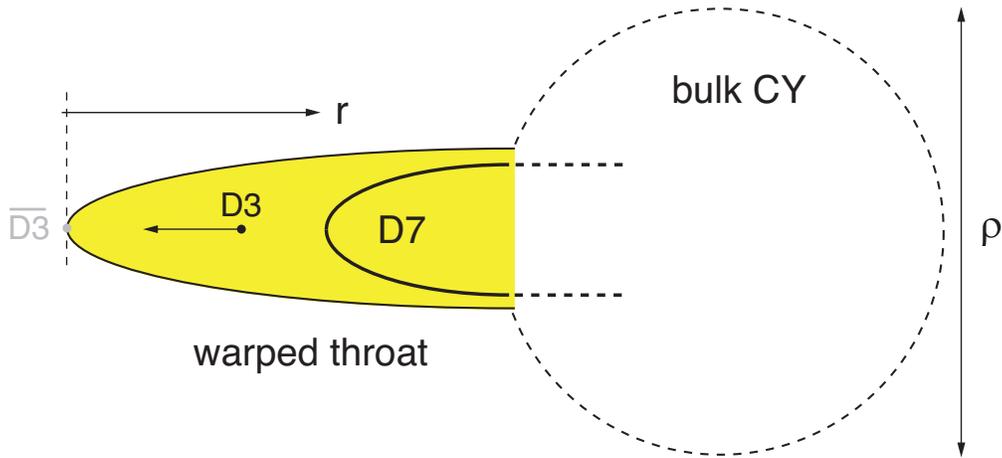}
    \caption{\small{Cartoon of an embedded stack of D7-branes wrapping a four-cycle $\Sigma_4$, and a mobile D3-brane, in a warped
    throat
region of a compact Calabi-Yau. In the scenario of \cite{KKLMMT}
the D3-brane feels a force from an anti-D3-brane at the tip of the
throat. Alternatively, in \cite{Dymarsky:2005xt} it was argued
that a D3-brane in the resolved warped deformed conifold
background feels a force even in the absence of an anti-D3-brane.
In this paper we consider an additional contribution to the
D3-brane potential, coming from nonperturbative effects on
D7-branes.}}
    \label{fig:throat}
\end{figure}

Let $\rho$ be the volume of a given four-cycle that admits a
nonperturbative effect.\footnote{In general, there are $h^{1,1}$
K\"ahler moduli $\rho_i$. For notational simplicity we limit our
discussion to a single K\"ahler modulus $\rho$, but point out that
our treatment straightforwardly generalizes to many moduli. The
identification of a holomorphic K\"ahler modulus, {\it{i.e.}} a
complex scalar belonging to a single chiral superfield, is
actually quite subtle. We address this important point in
\S\ref{subsec:rhoproblem}. At the present stage $\rho$ may simply
be taken to be the volume as defined in {\it{e.g.}} \cite{GKP}.}
The resulting superpotential is expected to be of the form
\cite{KKLT}
\begin{equation}
\label{equ:Wnp}
W_{{\rm{np}}}(\rho) = A(\chi,X)e^{-a \rho}\, .
\end{equation}
Here $a$ is a numerical constant and $A(\chi,X)$ is a
holomorphic function of the complex structure moduli $\chi \equiv
\{\chi_{1},\ldots, \chi_{h^{2,1}}\}$ and of the positions $X$ of
any D3-branes in the internal space.\footnote{Strictly speaking, there are three complex fields,
corresponding to the dimensionality of the internal space, but we
will refer to a single field for notational convenience.}
The functional form of
$A$ will depend on the particular four-cycle in question.

The prefactor $A(\chi,X)$ arises from a one-loop correction to the
nonperturbative superpotential.  For a Euclidean D3-brane
superpotential, $A(\chi,X)$ represents a one-loop determinant of
fluctuations around the instanton. In the case of D7-brane gauge
dynamics the prefactor comes from a threshold correction to the
gauge coupling on the D7-branes.

In the original KKLT proposal, the complex structure moduli
acquired moderately large masses from the fluxes, and no probe
D3-brane was present.  Thus, it was possible to ignore the
moduli-dependence of $A(\chi,X)$ and treat $A$ as a constant,
albeit an unknown one.  In the case of present interest (as in
\cite{KKLMMT}), the complex structure moduli are still massive
enough to be negligible, but there is at least one mobile D3-brane
in the compact space, so we must write $A=A(X)$.  (See
\cite{Ganor} for a very general argument that no prefactor $A$
can be independent of a D3-brane location $X$.)

The goal of this paper is to compute $A(X)$.  As we explained in
the introduction, this has already been achieved in certain
toroidal orientifolds \cite{BHK}, and the relevance of $A(X)$ for
brane inflation has also been recognized \cite{KKLMMT,BHK,refs}.
Here we will use a closed-string channel method for computing
$A(X)$, allowing us to study more general compactifications.  In
particular, we will give the first concrete results for $A(X)$ in
the warped throats relevant for many brane inflation models.

\subsection{D3-brane Potential After Volume Stabilization}
\label{subsec:enumeration}

 The $F$-term part of the supergravity potential is \beq
\label{equ:Fterm} V_F = e^{ \kappa_{4}^2{\cal K}} \left[{\cal
K}^{i \bar{j}} D_i W \overline{D_j W} - 3\kappa_{4}^2 |W|^2
\right]\, . \eeq DeWolfe and Giddings \cite{DeWolfe} showed that
the K\"ahler potential ${\cal K}$ in the presence of mobile
D3-branes is \beq \label{equ:Kahler} \kappa_4^2 {\cal K} = - 3
\log \left[\rho + \bar{\rho}-\gamma k(X,\bar{X})\right] \equiv -2
\log {\cal V}\, , \eeq where $k(X,\bar{X})$ is the K\"ahler
potential for the Calabi-Yau metric, {\it{i.e.}} the K\"ahler
potential on the putative moduli space of a D3-brane probe, ${\cal
V}$ is the physical volume of the internal space, and $\gamma$ is
a constant.\footnote{In \S\ref{subsec:rhoproblem} we will find
that $\gamma \equiv \frac{1}{3}\kappa_{4}^2 T_3$, where $T_3$ is
the D3-brane tension.} We address this volume-inflaton mixing in
more detail in \S\ref{subsec:rhoproblem}. For clarity we have
assumed here that there is only one K\"ahler modulus, but our
later analysis is more general.

The superpotential $W$ is the sum of a constant flux term
\cite{GVW} $W_{\rm flux}(\chi_\star) = \int G \wedge \Omega \equiv
W_0$ at fixed complex structure $\chi_\star$ and a term $W_{\rm
np}$~(\ref{equ:Wnp}) from nonperturbative effects, \beq
\label{equ:W} W = W_0 + A(X) e^{-a \rho}\, . \eeq

Equations (\ref{equ:Fterm}) to (\ref{equ:W}) imply three distinct
sources for corrections to the potential for D3-brane motion:

\begin{enumerate}
\item $m_{\cal K}$: The $X$-dependence of the K\"ahler potential
${\cal{K}}$ leads to a mass term familiar from the supergravity
eta problem.
\item $m_D$: Sources of $D$-term energy, if present, will scale with
the physical volume ${\cal V}$ and hence depend on the D3-brane
location.  This leads to a mass term for D3-brane displacements.
\item $m_A$: The prefactor $A(X)$ in the superpotential (\ref{equ:W})
leads to a mass term via the $F$-term potential
$(\ref{equ:Fterm})$.
\end{enumerate}

The masses $m_{\cal K}$ and $m_D$ were calculated explicitly in
\cite{KKLMMT} and shown to be of order the Hubble parameter $H$.
On the other hand, $m_A$ has been computed only for the toroidal
orientifolds of \cite{BHK}.  It has been suggested \cite{KKLMMT}
that there might exist non-generic configurations in which $m_A$
cancels against the other two terms. It is in these fine-tuned
situations that D3-brane motion could produce slow-roll inflation.
By computing $m_{A}$ explicitly, one can determine whether or not
this hope is realized \cite{PaperII}.

\section{Warped Volumes and the Superpotential}
\label{sec:backreaction}

\subsection{The Role of the Warped Volume}

The nonperturbative effects discussed in \S\ref{subsec:np} depend
exponentially on the warped volume of the associated four-cycle:
the warped volume governs the instanton action in the case of
Euclidean D3-branes, and the gauge coupling in the case of strong
gauge dynamics on D7-branes.  To see this, consider a warped
background with the line element \beq \label{equ:lineelt} \d s^2 =
G_{\mu\nu} \d x^{\mu} \d x^{\nu}+ G_{ij} \d Y^i \d Y^j \equiv
h^{-{1/2}}(Y) g_{\mu\nu} \d x^{\mu} \d x^{\nu} +
h^{{1/2}}(Y)g_{ij} \d Y^i \d Y^j \, , \eeq where $Y^i$ and
$g_{ij}$ are the coordinates and the unwarped metric on the
internal space, respectively, and $h(Y)$ is the warp factor.

The Yang-Mills coupling $g_7$ of the $7+1$ dimensional gauge
theory living on a stack of D7-branes is given by\footnote{In the
notation of \cite{Polchinski},  $g_{7}^2 = 2 g_{D7}^2$.} \beq
g_7^2 \equiv 2(2\pi)^5 g_s (\alpha')^2 \, .\eeq The action for
gauge fields on D7-branes that wrap a four-cycle $\Sigma_{4}$ is
\beq \label{equ:D7} S  = \frac{1}{2g_7^2} \int_{\Sigma_4} \d^4 \xi
\sqrt{g^{ind}} \, h(Y) \cdot \int \d^4 x \sqrt{g} \  g^{\mu
\alpha}  g^{\nu \beta}\, {\rm Tr}\, F_{\mu\nu}F_{\alpha \beta} \,
, \eeq where $\xi_{i}$ are coordinates on $\Sigma_4$ and $g^{ind}$
is the metric induced on $\Sigma_4$ from $g_{ij}$. A key point is
the appearance of a single power of $h(Y)$ \cite{Giddings}.
Defining the warped volume of $\Sigma_{4}$, \beq V_{\Sigma_4}^{w}
\equiv \int_{\Sigma_4} \d^4 \xi \sqrt{g^{ind}} \, h(Y)\, \eeq and
recalling the D3-brane tension \beq T_{3} \equiv \frac{1}{(2\pi)^3
g_s (\alpha')^2} \eeq we read off the gauge coupling of the
four-dimensional theory from (\ref{equ:D7}): \beq
\label{equ:coupling} \frac{1}{g^2} =
\frac{V_{\Sigma_4}^{w}}{g_7^2}= \frac{T_{3}
V_{\Sigma_4}^{w}}{8\pi^2}\, . \eeq

In ${\cal{N}}=1$ super-Yang-Mills theory, the Wilsonian gauge
coupling is the real part of a holomorphic function which receives
one-loop corrections, but no higher perturbative corrections
\cite{shifman,Amati,Seiberg}. The modulus of the gaugino
condensate superpotential in $SU(N_{D7})$ super-Yang-Mills with
ultraviolet cutoff $M_{\rm{UV}}$ is given by  \beq
\label{equ:gaugino} |W_{\rm np}| = {16\pi^2}M_{\rm{UV}}^3
\exp\Bigl(-\frac{1}{N_{D7}}\frac{8\pi^2}{g^2}\Bigr) \propto
\exp\Bigl(-\frac{T_{3} V_{\Sigma_4}^{w}}{N_{D7}} \Bigr)\, . \eeq
The mobile D3-brane adds a flavor to the $SU(N_{D7})$ gauge
theory, whose mass $m$ is a holomorphic function of the D3-brane
coordinates. In particular, the mass vanishes when the D3-brane
coincides with the D7-brane. In such a gauge theory, the
superpotential is proportional to $m^{1/N_{D7}}$
\cite{Intriligator}. Our explicit closed-string channel
calculations will confirm this form of the superpotential.

In the case that the nonperturbative effect comes from a Euclidean
D3-brane, the instanton action is \beq S = T_{3} \int_{\Sigma_4}
\d^4 \xi \sqrt{G^{ind}} =
 T_{3}\int_{\Sigma_4} \d^4 \xi \sqrt{g^{ind}}\, h(Y) \equiv
T_{3} V_{\Sigma_4}^{w}\, ,  \eeq so that, just as in
(\ref{equ:D7}), the action depends on a single power of $h(Y)$.
The modulus of the nonperturbative superpotential is then \beq
|W_{\rm np}| \propto \exp \Bigl( -T_{3} V_{\Sigma_4}^{w} \Bigr)\,
. \eeq

\subsection{Corrections to the Warped Volumes of Four-Cycles}
The displacement of a D3-brane in the compactification creates a
slight distortion $\delta h$ of the warped background, and hence
affects the warped volumes of four-cycles. The correction takes
the form
 \beq
 \label{equ:Delta}
\delta V_{\Sigma_4}^{w} \equiv \int_{\Sigma_4} \d^4 Y \sqrt{g^{
ind}(X;Y)}\, \delta h(X;Y)\, . \eeq By computing this change in
volume we will extract the dependence of the superpotential on the
D3-brane location $X$. In the non-compact throat approximation, we
will calculate $\delta V_{\Sigma_4}^{w}$ explicitly, and find that
it is the real part of a holomorphic function $\zeta
(X)$.\footnote{In the compact case, it is no longer true that
$\delta V_{\Sigma_4}^{w}$ is the real part of a holomorphic
function. This is related to the `rho problem' \cite{BHK}, and in
fact leads to a resolution of the problem, as we shall explain in
\S\ref{subsec:rhoproblem}.  The result is that in terms of an
appropriately-defined holomorphic K\"ahler modulus $\rho$
(\ref{equ:rhor}), the holomorphic correction to the gauge coupling
coincides with the holomorphic result of our non-compact
calculation.} Its imaginary part is determined by the integral of
the Ramond-Ramond four-form perturbation $\delta C_{4}$ over
$\Sigma_4$ (we will not compute this explicitly, but will be able
to deduce the result using the holomorphy of $\zeta (X)$).

The nonperturbative superpotential of the form (\ref{equ:Wnp}),
generated by the gaugino condensation, is then determined by \beq
\label{equ:aofX} A(X) = A_0
\exp\Bigl(-\frac{T_{3}~\zeta(X)}{N_{D7}}\Bigr)\, . \eeq We have
introduced an unimportant constant $A_0$ that depends on the
values at which the complex structure moduli are stabilized, but
is independent of the D3-brane position.  As remarked above,
computing (\ref{equ:aofX}) is equivalent to computing the
dependence of the threshold correction to the gauge coupling on
the mass $m$ of the flavor coming from strings that stretch from
the D7-branes to the D3-brane.

In the case of Euclidean D3-branes, the change in the instanton
action is proportional to the change in the warped four-cycle
volume.  Hence, the nonperturbative superpotential is of the form
(\ref{equ:Wnp}) with \beq A(X) = A_0 \exp\Bigl(-T_{3}~ \zeta(X)
\Bigr)\, . \eeq  In this case, computing (\ref{equ:Delta}) is
equivalent to computing the D3-brane dependence of an instanton
fluctuation determinant.

Finally, we can write a unified expression that applies to both
sources of nonperturbative effects: \beq \label{equ:A} A(X) = A_0
\exp\Bigl(-\frac{T_{3}~\zeta(X)}{n} \Bigr)\, , \eeq
where $n=N_{D7}$ for the case of gaugino condensation on D7-branes
and $n=1$ for the case of Euclidean D3-branes.

\section{D3-brane Backreaction}
\label{sec:GFM}

\subsection{The Green's Function Method}

A D3-brane located at some position $X$ in a six-dimensional space
with coordinates $Y$ acts as a point source for a perturbation
$\delta h$ of the geometry: \beq \label{equ:laplace}
 - \nabla_Y^2 \delta h(X;Y) = {\cal C}\,
\left[\frac{\delta^{(6)}(X-Y)}{\sqrt{g(Y)}} - \rho_{bg}(Y)\right]
\, . \eeq That is, the perturbation $\delta h$ is a Green's
function for the Laplace problem on the background of interest.
Here ${\cal C} \equiv 2 \kappa_{10}^2 T_{3} = (2\pi)^4 g_s
(\alpha')^2$ ensures the correct normalization of a single
D3-brane source term relative to the four-dimensional
Einstein-Hilbert action.  A consistent flux compactification
contains a background charge density $\rho_{bg}(Y)$ which
satisfies \beq \int {\rm d}^6 Y \sqrt{g}\, \rho_{bg}(Y) = 1 \,
\eeq to account for the Gauss's law constraint on the compact
space \cite{GKP}.

To solve (\ref{equ:laplace}), we first solve \beq
\label{equ:easylaplace} -\nabla_{Y'}^2 \Phi(Y;Y') = - \nabla_Y^2
\Phi(Y;Y') = \frac{\delta^{(6)}(Y-Y')}{\sqrt{g}}-\frac{1}{V_6} \,
, \eeq where $V_6 \equiv \int \d^6 Y \sqrt{g}$. The solution to
(\ref{equ:laplace}) is then \beq \label{equ:solvedlaplace} \delta
h(X;Y) = {\cal C}\, \Bigl[ \Phi(X;Y) - \int \d^6 Y' \sqrt{g}\,
\Phi(Y;Y') \rho_{bg}(Y') \Bigr] \, . \eeq  We note for later use
that \beq \label{equ:modulicharge} - \nabla_X^2 \delta h(X;Y) =
{\cal C}\, \left[{\delta^{(6)}(X-Y) \over \sqrt{g(X)}} -
\frac{1}{V_6} \right] \, . \eeq This relation is independent of
the form of the background charge $\rho_{bg}$.

To compute $A(X)$ from (\ref{equ:A}), we simply solve for the
Green's function $\delta h$ obeying (\ref{equ:laplace}) and then
integrate $\delta h$ over the four-cycle of interest, according to
(\ref{equ:Delta}).

\subsection{Comparison with the Open-String Approach}
\label{sec:BHK}

Let us show that this supergravity (closed-string channel) method
is consistent with the results of BHK \cite{BHK}, where the
correction to the gaugino condensate superpotential was derived
via a one-loop open-string computation.\footnote{Some analogous
pairs of closed-string and open-string computations exist in the
literature, {\it{e.g.}} \cite{banks}.}

The analysis of \cite{BHK} applied to configurations of D7-branes
and D3-branes on certain toroidal orientifolds, {\it{e.g.}} $T^2
\times T^4/\mathbb{Z}_2$.   We introduce a complex coordinate $X$
for the position of the D3-branes on $T^{2}$, as well as a complex
structure modulus $\tau$ for $T^{2}$, and without loss of
generality we set the volume of $T^{4}/\mathbb{Z}_2$ to unity. Let
us consider the case where all the D7-branes wrap
$T^{4}/\mathbb{Z}_2$ and sit at the origin $X=0$ in $T^{2}$.

The goal is to determine the dependence of the gauge coupling on
the position $X$ of a D3-brane.  (The location of the D3-brane in
the $T^4/\mathbb{Z}_2$ wrapped by the D7-branes is immaterial.)
For this purpose, we may omit terms computed in \cite{BHK} that
depend only on the complex structure and not on the D3-brane
location. Such terms will only affect the D3-brane potential by an
overall constant.

Then, the relevant terms from equation (44) of \cite{BHK}, in our
notation\footnote{After the replacement $X \to w$, our definitions
of the theta functions and torus coordinates correspond to those
of \cite{Polchinski}; our $X$ differs from the $A$ of \cite{BHK}
by a factor of $2\pi$.}, are \beq \label{equ:bhkresult} \delta
\Bigl(\frac{8\pi^2}{g^2}\Bigr) = \frac{1}{4\pi\,{\rm Im}(\tau)}
\Bigl[ {\rm{Im}}(X) \Bigr]^2 - \frac{1}{2} {\rm{ln}}
\left|\vartheta_{1}\left(\frac{X}{2\pi}\Bigl|\Bigr.\tau\right)\right|^2\,
. \eeq

Let us now compare (\ref{equ:bhkresult}) to the result of the
supergravity computation.  In principle, the prescription of
equation (\ref{equ:Delta}) is to integrate the Green's function on
a six-torus over the wrapped four-torus.  However, we notice that
this procedure of integration will reduce the six-dimensional
Laplace problem to the Laplace problem on the two-torus
parametrized by $X$, \beq \label{equ:torusgreens} - \nabla_{X}^2
\delta h(X;0) = {\cal C}\, \left[ \delta^{(2)}(X) -
\frac{1}{V_{T^2}} \right] \, , \eeq where $V_{T^2} =  8\pi^2\,{\rm
Im}(\tau)$.  The correction to the gauge coupling, in the
supergravity approach, is then proportional to $\delta h(X;0)$.
Solving (\ref{equ:torusgreens}) and using (\ref{equ:coupling}), we
get exactly (\ref{equ:bhkresult}). We conclude that our method
precisely reproduces the results of \cite{BHK}, at least for those
terms that directly enter the D3-brane potential.

\subsection{A Model for the Four-Cycles}
\label{sec:modelfourcycles}

The closed-string channel approach to calculating $A(X)$ is
well-defined for any given background, but further assumptions are
required when no complete metric for the compactification is
available. Fortunately, explicit metrics are available for many
non-compact Calabi-Yau spaces, and at the same time, the
associated warped throat regions are of particular interest for
inflationary phenomenology.  For a given warped throat geometry,
our approach is to compute the D3-brane backreaction on specific
four-cycles in the non-compact, asymptotically conical space.  We
will demonstrate that this gives an excellent approximation to the
backreaction in a compactification in which the same warped throat
is glued into a compact bulk. In particular, we will show in
\S\ref{subsec:modeldependent} that the physical effect in question
is localized in the throat, {\it{i.e.}} is determined primarily by
the shape of the four-cycle in the highly warped region. The model
therefore only depends on well-known data, such as the specific
warped metric and the embedding equations of the four-cycles, and
is insensitive to the unknown details of the unwarped bulk.  In
principle, our method can be extended to general compact models
for which metric data is available.

It still remains to identify the four-cycles responsible for
nonperturbative effects in this model of a warped throat attached
to a compact space.  Such a space will in general have many
K\"ahler moduli, and hence, assuming that stabilization is
possible at all, will have many contributions to the
nonperturbative superpotential.  The most relevant term, for the
purpose of determining the D3-brane potential, is the term
corresponding to the four-cycle closest to the D3-brane.  For a
D3-brane moving in the throat region, this is the four-cycle that
reaches farthest down the throat. In addition, the gauge theory
living on the corresponding D7-branes should be making an
important contribution to the superpotential.

The nonperturbative effects of interest are present only when the
four-cycle satisfies an appropriate topological condition
\cite{Witten}, which we will not discuss in detail.\footnote{These
rules can be changed in the presence of flux. For recent progress,
see {\it{e.g.}} \cite{ruleswithflux}.}  This topological condition
is, of course, related to the global properties of the four-cycle,
whereas the effect we compute is dominated by the part of the
four-cycle in the highly-warped throat region, and is insensitive
to details of the four-cycle in the unwarped region.  That is, our
methods are not sensitive to the distinction between four-cycles
that do admit nonperturbative effects, and those that do not. We
therefore propose to model the four-cycles producing
nonperturbative effects with four-cycles that are merely
supersymmetric, {\it{i.e.}} can be wrapped supersymmetrically by
D7-branes.  Many members of the latter class are not members of
the former, but as the shape of the cycle in the highly-warped
region is the only important quantity, we expect this distinction
to be unimportant.

We are therefore led to consider the backreaction of a D3-brane on
the volume of a stack of supersymmetric D7-branes wrapping a
four-cycle in a warped throat geometry. The simplest configuration
of this sort is a supersymmetric `flavor brane' embedding of
several D7-branes in a conifold \cite{KarchKatz,Ouyang,ACR}.

\section{Backreaction in Warped Conifold Geometries}
\label{sec:conifold} We now recall some relevant geometry.  The
singular conifold\footnote{The KS geometry \cite{KS} and its
generalizations \cite{Butti} are warped versions of the
{\it{deformed}} conifold, defined by $\sum_{i=1}^4 z_i^2 =
\varepsilon^2$.  When the D3-branes and D7-branes are sufficiently
far from the tip of the deformed conifold, it will suffice to
consider the simpler case of the warped singular conifold
constructed in \cite{KT}.} is a non-compact Calabi-Yau threefold
defined as the locus \beq \label{equ:conifoldconstraint}
\sum_{i=1}^4 z_i^2 = 0 \eeq in $\mathbb{C}^4$. After a linear
change of variables ($w_1= z_1+iz_2, w_2=z_1-iz_2$, {\it{etc}}.),
the constraint (\ref{equ:conifoldconstraint}) becomes \beq w_1 w_2
- w_3 w_4 = 0\, . \eeq The Calabi-Yau metric on the conifold is
\beq \label{equ:conifoldmetric} \d s_6^2 = \d r^2 + r^2 \d
s_{T^{1,1}}^2 \, . \eeq The base of the cone is the $T^{1,1}$
coset space $(SU(2)_A \times SU(2)_B)/U(1)_{R}$ whose metric in
angular coordinates $\theta_i \in [0,\pi], \phi_i \in [0,2 \pi],
\psi \in [0,4 \pi]$ is \beq \d s_{T^{1,1}}^2 = \frac{1}{9} \Bigl(
\d \psi + \sum_{i=1}^2 \cos \theta_i \, \d \phi_i \Bigr)^2 +
\frac{1}{6} \sum_{i=1}^2 \Bigl( \d \theta_i^2 + \sin^2 \theta_i \,
\d \phi_i^2 \Bigr) \, . \eeq

A stack of $N$ D3-branes placed at the singularity $w_i=0$
backreacts on the geometry, producing the ten-dimensional metric
\beq \label{warpedcone} \d s_{10}^2= h^{-1/2} (r) \d x_4^2 +
h^{1/2}(r) \d s_6^2 \ , \eeq where the warp factor is \beq
\label{confwarp} h(r)= \frac{27\pi g_s N (\alpha')^2}{4 r^4} \ .
\eeq This is the $AdS_5\times T^{1,1}$ background of type IIB
string theory, whose dual ${\cal N}=1$ supersymmetric conformal
gauge theory was constructed in \cite{KW}. The dual is an
$SU(N)\times SU(N)$ gauge theory coupled to bi-fundamental chiral
superfields $A_1, A_2, B_1, B_2$, each having $R$-charge $1/2$.
Under the $SU(2)_A\times SU(2)_B$ global symmetry, the superfields
transform as doublets. If we further add $M$ D5-branes wrapped
over the two-cycle inside $T^{1,1}$, then the gauge group changes
to $SU(N+M)\times SU(N)$, giving a cascading gauge theory
\cite{KT,KS}. The metric remains of the form (\ref{warpedcone}),
but the warp factor is modified to \cite{KT,Herzog:2002ih} \be
\label{nonharm} h(r) = \frac{27 \pi (\alpha')^2}{4r^4}\Bigl[g_s N
+ b (g_s M)^2 \log \Bigl(\frac{r}{r_0}\Bigr) +
 \frac{1}{4} b (g_s M)^2\Bigr]\, , \ee with $b \equiv
 \frac{3}{2\pi}$, and
$r_0\sim \varepsilon^{2/3} e^{2\pi N/(3 g_s M)}$. If an extra
D3-brane is added at small $r$, it produces a small change of the
warp factor, $\delta h = \frac{27 \pi g_s (\alpha')^2}{4 r^4} +
{\cal O}(r^{-11/2})$. A precise determination of $\delta h$ on the
conifold, using the Green's function method, is one of our goals
in this paper. As discussed above, this needs to be integrated
over a supersymmetric four-cycle.

\subsection{Supersymmetric Four-Cycles in the Conifold}
\label{sec:ACR}

The complex coordinates $w_i$ can be
related to the real coordinates $(r, \theta_i,\phi_i,\psi)$ via
\begin{eqnarray}
w_1 &=& r^{3/2} e^{\frac{i}{2}(\psi-\phi_1-\phi_2)} \sin \frac{\theta_1}{2} \sin \frac{\theta_2}{2}\, ,\\
w_2 &=& r^{3/2} e^{\frac{i}{2}(\psi+\phi_1+\phi_2)} \cos \frac{\theta_1}{2} \cos \frac{\theta_2}{2}\, ,\\
w_3 &=& r^{3/2} e^{\frac{i}{2}(\psi+\phi_1-\phi_2)} \cos \frac{\theta_1}{2} \sin \frac{\theta_2}{2}\, ,\\
w_4 &=& r^{3/2} e^{\frac{i}{2}(\psi-\phi_1+\phi_2)} \sin \frac{\theta_1}{2} \cos \frac{\theta_2}{2}\, .
\end{eqnarray}
It was shown in \cite{ACR} that the following holomorphic
four-cycles admit supersymmetric D7-branes:\footnote{This is not
an exhaustive list: another holomorphic embedding was used in
\cite{Kuperstein}.} \beq \label{equ:Embedding} f(w_i) \equiv
\prod_{i=1}^4 w_i^{p_i} - \mu^P = 0 \, . \eeq
Here $p_i \in \mathbb{Z}$, $P \equiv \sum_{i=1}^4 p_i$, and $\mu
\in \mathbb{C}$ are constants defining the embedding of the
D7-branes. In real coordinates the embedding condition
(\ref{equ:Embedding}) becomes
\begin{eqnarray} \label{equ:psi}
\psi(\phi_1,\phi_2) &=& n_1 \phi_1 + n_2 \phi_2 + \psi_s\, ,\\
r(\theta_1,\theta_2) &=& r_{\rm min} \left[x^{1+n_1} (1-x)^{1-n_1} y^{1+n_2} (1-y)^{1-n_2}\right]^{-1/6}\, ,
\end{eqnarray}
where
\bea
r_{\rm min}^{3/2} &\equiv& |\mu|\, ,\\
\frac{1}{2} \psi_s &\equiv& {\rm arg}(\mu) + \frac{2 \pi s}{P}\, ,
\quad s \in \{0,1,\dots,P-1\}\, . \eea We have defined the
coordinates \beq \label{equ:xandy} x \equiv \sin^2
\frac{\theta_1}{2}\, , \quad y \equiv \sin^2 \frac{\theta_2}{2}
\eeq and the rational winding numbers \beq \label{equ:wind} n_1
\equiv \frac{p_1-p_2-p_3+p_4}{P}\, , \quad n_2 \equiv
\frac{p_1-p_2+p_3-p_4}{P}\, . \eeq

To compute the integral over the four-cycle we will need the
volume form on the wrapped D7-brane, which is \beq
\label{equ:induced} {\rm d} \theta_1 {\rm d} \theta_2 {\rm d}
\phi_1 {\rm d} \phi_2\, \sqrt{g^{ind}} =
\frac{V_{T^{1,1}}}{16\pi^3} \, r^4 \, {\cal G}(x,y)\, {\rm d} x
{\rm d} y {\rm d} \phi_1 {\rm d} \phi_2 \, , \eeq where
\begin{eqnarray}
 {\cal G}(x,y) &\equiv& \frac{(1+n_1)^2}{2} \frac{1}{x(1-x)} - 2 n_1 \frac{1}{1-x} \nonumber \\
&+& \frac{(1+n_2)^2}{2} \frac{1}{y(1-y)} - 2 n_2 \frac{1}{1-y} -1\, .
\label{equ:induced2}
\end{eqnarray}
In (\ref{equ:induced}) we defined the volume of $T^{1,1}$ \beq
V_{T^{1,1}} \equiv \int \d^5 \Psi \sqrt{g_{T^{1,1}}} =\frac{16
\pi^3}{27}\, , \eeq with $\Psi$ standing for all five angular
coordinates on $T^{1,1}$.

For applications to brane inflation, we are interested in
four-cycles that do not reach the tip of the conifold ($|n_i| \le
1$). Two simple special cases of (\ref{equ:Embedding}) have this
property:
\begin{itemize}
\item {\it Ouyang embedding} \cite{Ouyang}:
$$w_1 = \mu\, .$$
\item {\it Karch-Katz embedding} \cite{KarchKatz}:
$$ w_1 w_2 = \mu^2\, . $$
\end{itemize}

Analogous supersymmetric four-cycles are known
\cite{ypqsusybranes} in some more complicated asymptotically
conical spaces, such as cones over $Y^{p,q}$ manifolds. We will
consider this case in \S\ref{subsec:ypq} and in Appendix
\ref{sec:ypq}.

\subsection{Relation to the Dual Gauge Theory Computation}
\label{sec:dual}

The calculation of $\delta h$ and its integration over a
holomorphic four-cycle is not sensitive to the background warp
factor. Let us discuss a gauge theory interpretation of the
calculation when we choose the background warp factor from
(\ref{confwarp}), {\it{i.e.}} we ignore the effect of $M$ wrapped
D5-branes. Here the gauge theory is exactly conformal, and we may
invoke the AdS/CFT correspondence to give a simple meaning to the
multipole expansion of $\delta h$, \beq \label{newwarp} \delta h=
\frac{27 \pi g_s (\alpha')^2}{4 r^4 }\left [1+ \sum_i \frac{c_i
f_i (\theta_1, \theta_2, \phi_1, \phi_2, \psi)}{r^{\Delta_i}}
\right ] \ . \eeq In the dual gauge theory, the $c_i$ are
proportional to the expectation values of gauge-invariant
operators ${\cal O}_i$ determined by the position of the D3-brane
\cite{KW2}. Among these operators a special role is played by the
chiral operators of $R$-charge $k$, ${\rm Tr} [A_{\alpha_1}
B_{\dot{\beta}_1} A_{\alpha_2} B_{\dot{\beta}_2}\ldots
A_{\alpha_k} B_{\dot{\beta}_k}]$, symmetric in both the dotted and
the undotted indices. These operators have exact dimensions
$\Delta_i^{\rm chiral} = 3k/2$ and transform as $(k/2, k/2)$ under
the $SU(2)_A\times SU(2)_B$ symmetry. In addition to these
operators, many non-chiral operators, whose dimensions $\Delta_i$
are not quantized \cite{Gubser:1998vd}, acquire expectation values
and therefore affect the multipole expansion of the warp factor.
But remarkably, {\it all} these non-chiral contributions vanish
upon integration over a holomorphic four-cycle. Therefore, the
contributing terms in $\delta h$ have the simple form \cite{KW2}
\beq \label{trulynewwarp} \delta h_{\rm chiral}= \frac{27 \pi g_s
(\alpha')^2}{4 r^4 }\left [1+ \sum_{k=1}^\infty
\frac{\left(f_{a_1\dots a_k} \widehat z_{a_1\dots
a_k}+c.c.\right)}{r^{3k/2}}\right] \ , \eeq where $f_{a_1\dots
a_k}\sim \bar\epsilon_{a_1} \bar\epsilon_{a_2} \ldots
\bar\epsilon_{a_k}$ for a D3-brane positioned at $z_{a}
=\epsilon_{a}$. Above, $\widehat z_{a_1\dots a_k}$ are the
normalized spherical harmonics on $T^{1,1}$ that transform as
$(k/2, k/2)$ under the $SU(2)_A\times SU(2)_B$. The normalization
factors are defined in Appendix \ref{sec:GF}.

The leading term in (\ref{newwarp}), which falls off as $1/r^4$,
gives a logarithmic divergence at large $r$ when integrated over a
four-cycle. We note that this term does not appear if we define
$\delta h$ as the solution of (\ref{equ:laplace}) with $\sqrt{g}\,
\rho_{bg}(Y) = \delta^{(6)}(Y-X_0)$. This corresponds to
evaluating the change in the warp factor, $\delta h$, created by
moving the D3-brane to $X$ from some reference point $X_0$.  If we
choose the reference point $X_0$ to be at the tip of the cone,
$r=0$, then (\ref{newwarp}) is modified to \beq
\label{othernewwarp} \delta h= \frac{27 \pi g_s (\alpha')^2}{4 r^4
}\left [ \sum_i \frac{c_i f_i (\theta_1, \theta_2, \phi_1, \phi_2,
\psi)}{r^{\Delta_i}}\right ] \ . \eeq An advantage of this
definition is that now there is a precise correspondence between
our calculation and the expectation values of operators in the
dual gauge theory.

\subsection{Results for the Conifold}
\label{subsec:results}

We are now ready to compute the D3-brane-dependent correction to
the warped volume of a supersymmetric four-cycle in the conifold.
Using the Green's function on the singular conifold
(\ref{equ:conifoldgreens}), which we derive in Appendix
\ref{sec:GF}, and the explicit form of the induced metric
$\sqrt{g^{ind}}$ (\ref{equ:induced}), we carry out integration
term by term and find that most terms in (\ref{equ:Delta}) do not
contribute.  We relegate the details of this computation to
Appendix \ref{sec:Calc}.  As we demonstrate in Appendix
\ref{sec:Calc}, the terms that do not cancel are precisely those
corresponding to (anti)chiral deformations of the dual gauge
theory.

Integrating (\ref{othernewwarp}) term by term as prescribed in
(\ref{equ:Delta}), we find that the final result for a general
embedding (\ref{equ:Embedding}) is \beq \label{equ:finalDf} T_3\,
\delta V_{\Sigma_4}^w = T_3 \, {\rm Re } \left(\zeta(w_i)\right) =
- {\rm Re} \left(  \log \left[\frac{ \mu^P - \prod_{i=1}^4
w_i^{p_i} }{\mu^P} \right] \right) \, ,
 \eeq
 so that
 \beq
 \label{equ:finalA}
 A =A_0 \left( \frac{  \mu^{P} -  \prod_{i=1}^4 w_i^{p_i}
}{\mu^P} \right)^{1/n}\, . \eeq Comparing to
(\ref{equ:Embedding}), we see that $A$ is proportional to a power
of the holomorphic equation that specifies the embedding.  For $n=N_{D7}$
coincident D7-branes, this power is $1/n$.  This behavior agrees
with the results of \cite{Ganor}; note in particular that when
$n=1$, (\ref{equ:finalA}) has a simple zero everywhere on the
four-cycle, as required by \cite{Ganor}.

Finally, let us specialize to the two cases of particular
interest, the Ouyang \cite{Ouyang} and Karch-Katz \cite{KarchKatz}
embeddings in which the four-cycle does not reach all the way to
the tip of the throat. For the Ouyang embedding we find \beq
A(w_1) = A_0 \left( \frac{\mu - w_1}{\mu} \right)^{1/n}\, , \eeq
whereas for the Karch-Katz embedding we have \beq A(w_1,w_2) = A_0
\left( \frac{\mu^2 - w_1 w_2}{\mu^2}\right)^{1/n}\, . \eeq

\subsection{Results for $Y^{p,q}$ Cones}
\label{subsec:ypq}

Recently, a new infinite class of Sasaki-Einstein manifolds
$Y^{p,q}$ of topology $S^2\times S^3$ was discovered
\cite{Gauntlett2, Gauntlett}. The ${\cal N}=1$ superconformal
gauge theories dual to $AdS_5\times Y^{p,q}$ were constructed in
\cite{Benvenuti}. These quiver theories, which live on $N$
D3-branes at the apex of the Calabi-Yau cone over $Y^{p,q}$, have
gauge groups $SU(N)^{2p}$, bifundamental matter, and marginal
superpotentials involving both cubic and quartic terms. Addition
of $M$ D5-branes wrapped over the $S^2$ at the apex produces a
class of cascading gauge theories whose warped cone duals were
constructed in \cite{HEK}. A D3-brane moving in such a throat
could also serve as a model of D-brane inflation
\cite{Dymarsky:2005xt}.

Having described the calculation for the singular conifold in some detail,
we now cite the results of an equivalent computation for cones
over $Y^{p,q}$ manifolds.
More details can be found in Appendix \ref{sec:ypq}.

Supersymmetric four-cycles in $Y^{p,q}$ cones are defined by the
following embedding condition \cite{ypqsusybranes} \beq f(w_i)
\equiv \prod_{i=1}^3 w_i^{p_i} - \mu^{2 p_3} = 0 \, , \eeq
where the complex coordinates $w_i$ are defined in Appendix
\ref{sec:ypq}. Integration of the Green's function over the
four-cycle leads to the following result for the perturbation to
the warped volume \beq \label{equ:finalDfY} T_3\, \delta
V_{\Sigma_4}^w = T_3 \, {\rm Re } \left(\zeta(w_i)\right) = - {\rm
Re} \left(  \log \left[\frac{ \mu^{2 p_3} - \prod_{i=1}^3
w_i^{p_i} }{\mu^{2 p_3}} \right] \right) \, ,
 \eeq
 so that
 \beq
 \label{equ:finalAY}
 A =A_0 \left( \frac{  \mu^{2 p_3} -  \prod_{i=1}^3 w_i^{p_i}
}{\mu^{2 p_3}} \right)^{1/n}\, . \eeq

\subsection{General Compactifications}

The arguments in \cite{Ganor}, which were based on studying the
change in the theta angle as one moves the D3-brane around the
D7-branes, indicate that the correction is a section of a bundle
called the `divisor bundle'. This section has a zero at the
location of the D7-branes.  The correction has to live in a
non-trivial bundle since a holomorphic function on a compact space
would be a constant. In the non-compact examples we considered
above we can work in only one coordinate patch and obtain the
correction as a simple function, the function characterizing the
embedding. Strictly speaking, the arguments in \cite{Ganor} were
made for the case that the superpotential is generated by wrapped
D3-instantons. But the same arguments can be used to compute the
correction for the gauge coupling on D7-branes.

In summary, we have explicitly computed the modulus of $A$, and
found a result in perfect agreement with the analysis of the phase
of $A$ in \cite{Ganor}.  One has a general answer of the form \beq
\label{equ:generalanswer} A(w_i) = A_0~\Bigl( f(w_i) \Bigr)^{ 1/n}
\, , \eeq where $f$ is a section of the divisor bundle and
$f(w_i)=0$ specifies the location of the D7-branes.

\section{Compactification Effects}
\label{sec:compactification}
\subsection{Holomorphy of the Gauge Coupling}
\label{subsec:rhoproblem}

In compactifications with mobile D3-branes, the identification of
holomorphic K\"ahler moduli and holomorphic gauge couplings is
quite subtle. This has become known as the `rho problem'
\cite{BHK,Giddings}.\footnote{Similar issues were discussed in
\cite{Wittencomp}.} Let us recall the difficulty.  In the internal
metric $g_{ij}$ appearing in (\ref{equ:lineelt}), we can identify
the breathing mode of the compact space via \beq
\label{equ:breathe} g_{ij} \equiv \tilde{g}_{ij} e^{2u}\, , \eeq
where $\tilde{g}_{ij}$ is a fiducial metric. In the following, all
quantities computed from $\tilde{g}_{ij}$ will be denoted by a
tilde. The Born-Infeld kinetic term for a D3-brane, expressed in
Einstein frame and in terms of complex coordinates $X,\bar{X}$ on
the brane configuration space, is then \beq \label{equ:dbi}
S_{kin} = -T_3 \int \d^4 x
\sqrt{g}e^{-4u}\partial_{\mu}X^{i}\partial^{\mu}\bar{X}^{\bar{j}}\tilde{g}_{i\bar{j}}\,
. \eeq DeWolfe and Giddings argued in \cite{DeWolfe} that to
reproduce this volume scaling, as well as the known no-scale,
sequestered property of the D3-brane action in this background,
the K\"ahler potential must take the form \beq \label{equ:ronly}
{\kappa_{4}^2} {\cal K} = - 3 \log{e^{4u}}\, ,\eeq with the
crucial additional requirement that \beq
\partial_{i}\partial_{\bar{j}} e^{4u} \propto
\tilde{g}_{i\bar{j}}\, , \eeq so that $e^{4u}$ contains a term
proportional to the K\"ahler potential $k(X,\bar{X})$ for the
fiducial Calabi-Yau metric.  Comparing (\ref{equ:dbi}) to the
kinetic term derived from (\ref{equ:ronly}), we find in fact \beq
\label{equ:laplacek}
\partial_{i}\partial_{\bar{j}} e^{4u} =
-\left(\frac{\kappa_{4}^2 T_3}{3}\right)k_{, i\bar{j}}
 \, . \eeq
We can now define the holomorphic volume modulus $\rho$ as
follows. The real part of $\rho$ is given by \beq \label{equ:rhor}
\rho+\bar{\rho} \equiv e^{4u} + \left(\frac{\kappa_{4}^2
T_3}{3}\right)k(X,\bar{X}) \eeq and the imaginary part is the
axion from the Ramond-Ramond four-form potential.  As explained in
\cite{KKLMMT}, this is consistent with the fact that the axion
moduli space is a circle that is non-trivially fibered over the
D3-brane moduli space.

Next, the gauge coupling on a D7-brane is easily seen to be
proportional to the breathing mode of the metric, $e^{4u} \equiv
\rho+\bar{\rho}-\left(\frac{1}{3}{\kappa_{4}^2}T_3\right)k(X,\bar{X})$,
which is {\it{not}} the real part of holomorphic function on the
brane moduli space. However, supersymmetry requires that the gauge
kinetic function is a holomorphic function of the moduli.  This
conflict is the rho problem.

We can trace this problem to an incomplete inclusion of the
backreaction due to the D3-brane.  Through (\ref{equ:rhor}), the
physical volume modulus $e^{4u}$ has been allowed to depend on the
D3-brane position. That is, the difference between the holomorphic
modulus $\rho$ and the physical modulus $e^{4u}$ is affected by
the D3-brane position.  This was necessary in order to recover the
known properties of the brane/volume moduli space.  Notice from
(\ref{equ:rhor}) that the strength of this open-closed mixing is
controlled by ${\kappa_{4}^2}T_3$, and so is manifestly a
consequence of D3-brane backreaction in the compact space.
However, as we explained in \S\ref{sec:backreaction}, the warp
factor $h$ also depends on the D3-brane position, again via
backreaction.  To include the effects of the brane on the
breathing mode, but not on the warp factor, is not consistent.\footnote{Let us point out that this is precisely the closed-string dual of
the resolution found in \cite{BHK}: careful inclusion of the
open-string one-loop corrections to the gauge coupling resolved
the rho problem.  In that language, the initial inconsistency was
the inclusion of only some of the one-loop effects.}
One
might expect that consideration of the correction $\delta h$ to
the warp factor would restore holomorphy and resolve the rho
problem. We will now see that this is indeed the case.

What we find is that the uncorrected warped volume
$(V_{\Sigma_4}^w)_0$, as well as the correction $\delta
V_{\Sigma_4}^w$, are both non-holomorphic, but their
non-holomorphic pieces precisely cancel, so that the
{\it{corrected}} warped volume $V_{\Sigma_4}^{w}$ is the real part
of a holomorphic function of the moduli $\rho$ and $X$.

First, we separate the constant, zero-mode, piece of the warp
factor: \beq \label{equ:hsplit} h(X;Y) = h_0 + \delta h(X;Y) \, .
\eeq
By definition $\delta h(X;Y)$ integrates to zero over the compact
manifold, \beq \int \d^6 Y \sqrt{g(Y)} \, \delta h(X;Y) = 0 \, .
\eeq
This implies that the factor of the volume that appears in the
four-dimensional Newton constant is unaffected by $\delta h$.
Thus we have $\kappa_4^{-2} = \kappa_{10}^{-2} h_0 \tilde V_6 $.
 We
define the uncorrected warped volume via \beq \label{equ:uncorrgc}
(V_{\Sigma_4}^w)_0 \equiv \int_{\Sigma_4} \d^4 \xi \sqrt{g^{ind}}
\, h_0
 = e^{4u(X,\bar{X})} \, h_0 \, \tilde{V}_{\Sigma_4} \, . \eeq This is
non-holomorphic because of the prefactor $e^{4u(X,\bar{X})}$.  In
particular, using (\ref{equ:rhor}), we have \beq
\label{equ:constpart} (V_{\Sigma_4}^w)_0 = -\left(\frac{\kappa_4^2
T_3}{3}\right)  \tilde{V}_{\Sigma_4} \, h_0 \, k(X,\bar{X}) +
[{\rm hol.} + {\rm antihol.}]\, .
 \eeq

We next consider $\delta h$.  When the D3-brane is not coincident
with the four-cycle of interest, we find from
(\ref{equ:modulicharge}) that $\delta h$ obeys \beq
\label{equ:nablax} \nabla_{X}^2 \delta h(X;Y) =
 \frac{\cal{C}}{V_6}\,  \eeq
where ${\cal C} \equiv 2 \kappa_{10}^2 T_3 = 2 \kappa_4^2 T_3 h_0
\tilde{V}_6$.
Hence, $\delta h$ is {\it{not}} the real part of a
holomorphic function of $X$.  The source of the deviation from
holomorphy is the term $\frac{1}{V_6}$ in
(\ref{equ:modulicharge}).  Although this term is superficially
similar to a constant background charge density, it is independent
of the density $\rho_{bg}(Y)$ of physical D3-brane charge in the
internal space, which has coordinates $Y$.  Instead,
$\frac{1}{V_6}$ may be thought of as a `background charge' on the
D3-brane moduli space, which has coordinates $X$.  From this
perspective, it is the Gauss's law constraint on the D3-brane
moduli space that forces $\delta h$ to be non-holomorphic.

In complex coordinates, using the metric $\tilde{g}$, and noting
that $\tilde{V}_{6}=V_{6}\,e^{-6u}$, (\ref{equ:nablax}) may be
written as \beq
\tilde{g}^{i\bar{j}}\partial_{i}\partial_{\bar{j}}\delta h =
\kappa_4^2 T_3 \, h_0 \, e^{-4u} \, ,\eeq where because the
compact space is K\"ahler, we can write the Laplacian using
partial derivatives. It follows that \beq \label{equ:delisnonhol}
\delta h = \left(\frac{\kappa_4^2 T_3 }{3}\right) h_0 \, e^{-4u}
k(X,\bar{X}) + [{\rm hol.} + {\rm antihol.}]\, . \eeq The omitted
holomorphic and antiholomorphic terms are precisely those that we
computed in the preceding sections. Furthermore, recalling the
definition (\ref{equ:Delta}), we have \beq
\label{equ:integratednonhol} \delta V_{\Sigma_4}^w =
\left(\frac{\kappa_4^2 T_3 }{3}\right) h_0 \,
\tilde{V}_{\Sigma_4}k(X,\bar{X}) + [\zeta(X) +
\overline{\zeta(X)}]\, . \eeq The non-holomorphic first term in
(\ref{equ:integratednonhol}) precisely cancels the non-holomorphic
term in $(V_{\Sigma_4}^{w})_0$ (\ref{equ:constpart}), so that \beq
\label{equ:rhosolved} V_{\Sigma_4}^{w} = (V_{\Sigma_4}^{w})_0 +
\delta V_{\Sigma_4}^w = \tilde{V}_{\Sigma_4} \, h_0 \, (\rho +
\bar{\rho})+ [\zeta(X)+\overline{\zeta(X)}]  \,  .\eeq We conclude
that $V_{\Sigma_4}^{w}$ can be the real part of a holomorphic
function.\footnote{Strictly speaking, we have shown only that
$V_{\Sigma_4}^{w}$ is in the kernel of the Laplacian; the r.h.s.
of (\ref{equ:delisnonhol}) and (\ref{equ:rhosolved}) could in
principle contain extra terms that are annihilated by the
Laplacian but are not the real parts of holomorphic functions.
However, the obstruction to holomorphy presented by $k(X,\bar{X})$
has disappeared, and we expect no further obstructions.}

To summarize, we have seen that the background charge term in
(\ref{equ:modulicharge}), which was required by a constraint
analogous to Gauss's law on the D3-brane moduli space, causes
$\delta V_{\Sigma_4}^w$ to have a non-holomorphic term
proportional to $k(X,\bar{X})$. Furthermore, the DeWolfe-Giddings
K\"ahler potential produces a well-known non-holomorphic term,
also proportional to $k(X,\bar{X})$, in the uncorrected warped
volume $(V_{\Sigma_4}^w)_0$.  We found that these two terms
precisely cancel, so that the total warped volume $V_{\Sigma_4}^w
= (V_{\Sigma_4}^{w})_0 + \delta V_{\Sigma_4}^w$ can be
holomorphic. Thus, the corrected gauge coupling on D7-branes, and
the corrected Euclidean D3-brane action, are
holomorphic.\footnote{To complete the identification of the
holomorphic variable, we note that the constant $a$ appearing in
(\ref{equ:Wnp}) is $a \equiv 2 T_3 \tilde{V}_{\Sigma_4} h_0/n$.
The resulting dependence on $g_s$ could be absorbed by a
redefinition of $\rho$, as in \cite{KKLT}.}

Note that, as a consequence of this discussion, the holomorphic
part of the correction to the volume changes under K\"ahler
transformations of $k(X, \bar X)$. This implies that the
correction is in a bundle whose field strength is proportional to
the K\"ahler form.

\subsection{Model-Dependent Effects from the Bulk}
\label{subsec:modeldependent}

In \S\ref{subsec:enumeration}, we listed three contributions to
the potential for D3-brane motion.  The first two were given
explicitly in \cite{KKLMMT}, and we have computed the third.  It
is now important to ask whether this is an exhaustive list: in
other words, might there be further effects that generate D3-brane
mass terms of order $H$?  In particular, could coupling of the
throat to a compact bulk generate corrections to our results, and
hence adjust the brane potential?

First, let us justify our approach of using noncompact warped
throats to model D3-brane potentials in compact spaces with finite
warped throat regions.  The idea is that the effect of the
D3-brane on a four-cycle is localized in that portion of the
four-cycle that is deepest in the throat.  Comparing
(\ref{equ:induced}) to (\ref{othernewwarp}), we see that all
corrections to the warped volume scale inversely with $r$, and are
therefore supported in the infrared region of the throat.  Hence,
as anticipated in \S\ref{sec:modelfourcycles}, the effects of
interest are automatically concentrated in the well-understood
region of high warping, far from the model-dependent region where
the throat is glued into the rest of the compact space.  This is
true even though a typical four-cycle will have most of its volume
in the bulk, outside the highly warped region.  The perturbation
due to the D3-brane already falls off faster than $r^{-4}$ in the
throat, where the measure factor is $r^4$, and in the bulk the
perturbation will diminish even more rapidly. Except in remarkable
cases, the diminution of the perturbation will continue to
dominate the growth of the measure factor. A similar argument
reinforces our assertion that the dominant effect on a D3-brane
comes from whichever wrapped brane descends farthest into the
throat.

We conclude that the effects of the gluing region, where the
throat meets the bulk, and of the bulk itself, produce negligible
corrections to the terms we have computed.  Fortunately, the
leading effects are concentrated in the highly warped region,
where one has access to explicit metrics and can do complete
computations.

We have now given a complete account of the nonperturbative
superpotential.  However, the K\"ahler potential is not protected
against perturbative corrections, which could conceivably
contribute to the low-energy potential for D3-brane motion.
Explicit results are not available for general compact spaces
(see, however, \cite{Hebecker,BHK2}); here we will simply argue
that these corrections can be made negligible. Recall that the
DeWolfe-Giddings K\"ahler potential provides a mixing between the
volume and the D3-brane position that generates brane mass terms
of order $H$.  Any {\it{further}} corrections to the K\"ahler
potential, whether from string loops or sigma-model loops, will be
subleading in the large-volume, weak-coupling limit, and will
therefore generically give mass terms that are small compared to
$H$.  In addition, the results of \cite{fkt} give some constraints
on $\alpha'$ corrections to warped throat geometries.  We leave a
systematic study of this question for the future.

\section{Implications and Conclusion}
\label{sec:conclusion}

We have used a supergravity approach (see also \cite{Giddings}) to
study the D3-brane corrections to the nonperturbative
superpotential induced by D7-branes or Euclidean D3-branes
wrapping four-cycles of a compactification.  This has been a key,
unknown element of the potential governing D3-brane motion in such
a compactification.  We integrated the perturbation to the
background warping due to the D3-brane over the wrapped
four-cycle.  The resulting position-dependent correction to the
warped four-cycle volume modifies the strength of the
nonperturbative effect, which in turn implies a force on the
D3-brane.  This computation is the closed-string channel dual of
the threshold correction computation of \cite{BHK}, and we showed
that the closed-string method efficiently reproduces the results
of \cite{BHK}.

We then investigated the D3-brane potential in explicit warped
throat backgrounds with embedded wrapped branes. We showed that
for holomorphic embeddings, only those deformations corresponding
to (anti)chiral operators in the dual gauge theory contribute to
correcting the superpotential. This led to a strikingly simple
result: the superpotential correction is given by the embedding
condition for the wrapped brane, in accord with \cite{Ganor}.

An important application of our results is to cosmological models
with moving D3-branes, particularly warped brane inflation models
\cite{KKLMMT,otherstringinflation,Tye,Dymarsky:2005xt}. It is
well-known that these models suffer from an eta problem and hence
produce substantial inflation only if the inflaton mass term is
fine-tuned to fall in a certain range. Our result determines a
`missing' contribution to the inflaton potential that was
discussed in \cite{KKLMMT}, but was not computed there. Equipped
with this contribution, one can quantify the fine-tuning in warped
brane inflation by considering specific choices of throat
geometries and of embedded wrapped branes, and determining whether
prolonged inflation occurs \cite{PaperII}.  This amounts to a
microscopically justified method for selecting points or regions
within the phenomenological parameter space described in
\cite{Tye}. This approach was initiated in \cite{BHK}, but the
open-string method used there does not readily extend beyond
toroidal orientifolds, and is especially difficult for warped
throats in flux compactifications. In contrast, our concrete
computations were performed in warped throat backgrounds, and thus
apply directly to warped brane inflation models, including
backgrounds with fluxes.

Our approach also led to a natural solution of the `rho problem',
{\it{i.e.}} the apparent non-holomorphy of the gauge coupling on
wrapped D7-branes in backgrounds with D3-branes.  This problem
arises from incomplete inclusion of D3-brane backreaction effects,
and in particular from omission of the correction to the warped
volume that we computed in this work.  We observed that the
correction is itself non-holomorphic, as a result of a Gauss's law
constraint on the D3-brane moduli space.  Moreover, the
non-holomorphic correction cancels precisely against the
non-holomorphic term in the uncorrected warped volume, leading to
a final gauge kinetic function that is holomorphic.

In closing, let us emphasize that the problem of fine-tuning in
D-brane inflation models has {\it{not}} disappeared, but can now
be made more explicit. A detailed analysis of this will be
presented in a future paper \cite{PaperII}.

\section*{Acknowledgments}
We thank Cliff Burgess, Oliver DeWolfe, Alan Guth, Michael Haack,
Shamit Kachru, Renata Kallosh, Lev Kofman, John McGreevy, Gregory
Moore, Andrew Neitzke, Joseph Polchinski, Nathan Seiberg, Eva
Silverstein, Paul Steinhardt, Henry Tye, Herman Verlinde, and
Alexander Westphal for helpful discussions. DB thanks Marc
Kamionkowski and the Theoretical Astrophysics Group at Caltech for
their hospitality while some of this work was carried out. DB also
thanks the Institut d'Astrophysique de Paris for their hospitality
during the conference `Inflation+25'. LM thanks the Stanford
Institute for Theoretical Physics and the organizers of the ICTP
workshop `String Vacua and the Landscape' for their hospitality.
This research was supported in part by the United States
Department of Energy, under contracts DOE-FC02-92ER-40704 and
DE-FG02-90ER-40542, and by the National Science Foundation under
Grant No. PHY-0243680.  The research of AD is also supported in
part by Grant RFBR 04-02-16538, and Grant for Support of
Scientific Schools NSh-8004.2006.2.  Any opinions, findings, and
conclusions or recommendations expressed in this material are
those of the authors and do not necessarily reflect the views of
the National Science Foundation.
\newpage

\appendix

\section{Green's Functions on Conical Geometries}
\label{sec:GF}

\addtocontents{toc}{\SkipTocEntry}
\subsection{Green's Function on the Singular Conifold}
\label{subsec:con}

The D3-branes that we consider in this paper are point sources in
the six-dimensional internal space.  The backreaction they induce
on the background geometry can therefore be related to the Green's
functions for the Laplace problem on conical geometries (see
\S\ref{sec:GFM}) \beq \label{equ:GreenFct} -\nabla_X^2 G(X;X') =
\frac{\delta^{(6)}(X-X')}{\sqrt{g(X)}}\, . \eeq In the following
we present explicit results for the Green's function on the
singular conifold. In the large $r$-limit, far from the tip, the
Green's functions for the resolved and deformed conifold reduce to
those of the singular conifold.

In the singular conifold geometry (\ref{equ:conifoldmetric}), the
defining equation (\ref{equ:GreenFct}) for the Green's function
becomes \beq \label{equ:def} \frac{1}{r^5}
\frac{\partial}{\partial r} \Bigl( r^5 \frac{\partial }{\partial
r} G\Bigr) + \frac{1}{r^2}\nabla^2_\Psi G = -\frac{1}{r^5}
\delta(r-r') \delta_{T^{1,1}}(\Psi-\Psi')\, , \eeq where
$\nabla^2_\Psi$ and $\delta_{T^{1,1}}(\Psi-\Psi')$ are the
Laplacian and the normalized delta function on $T^{1,1}$,
respectively. $\Psi$ stands collectively for the five angular
coordinates of the base and $X \equiv (r,\Psi)$. An explicit
solution for the Green's function is obtained by a series
expansion of the form \beq G(X;X') = \sum_L Y^*_L(\Psi') Y_L(\Psi)
H_L(r;r')\, . \eeq The $Y_L$'s are eigenfunctions of the angular
Laplacian, \beq \nabla_\Psi^2 Y_L(\Psi) = - \Lambda_L Y_L(\Psi)\,
, \eeq where the multi-index $L$ represents the set of discrete
quantum numbers related to the symmetries of the base of the cone.
The angular eigenproblem is worked out in detail in
\S\ref{sec:eigen}. If the angular wavefunctions are normalized as
\beq \label{equ:normal} \int \d^5 \Psi \, \sqrt{g_{T^{1,1}}} \,
Y^*_{L}(\Psi) Y_{L'}(\Psi) = \delta_{L L'}\, , \eeq then \beq
\label{equ:DF} \sum_L Y_L^*(\Psi') Y_L(\Psi) =
\delta_{T^{1,1}}(\Psi-\Psi')\, , \eeq and equation (\ref{equ:def})
reduces to the radial equation \beq \label{equ:radial2}
\frac{1}{r^5} \frac{\partial}{\partial r} \Bigl( r^5
\frac{\partial}{\partial r} H_L \Bigr) - \frac{\Lambda_L}{r^2} H_L
= -\frac{1}{r^5} \delta(r-r')\, , \eeq whose solution away from
$r=r'$ is \beq H_L(r;r') = A_\pm(r') r^{c_L^\pm}\, , \quad \quad
c_L^\pm \equiv -2 \pm \sqrt{\Lambda_L+4}\, . \eeq The constants
$A_\pm$ are uniquely determined by integrating equation
(\ref{equ:radial2}) across $r=r'$. The Green's function on the
singular conifold is \beq \label{equ:conifoldgreens} G(X;X') =
\sum_L \frac{1}{2 \sqrt{\Lambda_L+4}}\times Y_L^*(\Psi') Y_L(\Psi)
\times
\begin{cases}
\frac{1}{r'^4} \Bigl(\frac{r}{r'}\Bigr)^{c_L^+} & r \le r'\, , \cr
\frac{1}{r^4} \Bigl(\frac{r'}{r}\Bigr)^{c_L^+} & r \ge r'\, ,
\end{cases}
\eeq
where the angular
eigenfunctions $Y_L(\Psi)$ are given explicitly in \S\ref{sec:eigen}.

\addtocontents{toc}{\SkipTocEntry}
\subsection{Eigenfunctions of the Laplacian on $T^{1,1}$}
\label{sec:eigen}

In this section we complete the Green's function on the singular conifold (\ref{equ:conifoldgreens}) by solving for the eigenfunctions of the Laplacian on $T^{1,1}$
\bea
\label{equ:eigen}
\nabla^2_\Psi Y_L &=& \frac{1}{\sqrt{g}} \partial_m (g^{mn} \sqrt{g} \partial_n Y_L) = (6 \nabla^2_1 + 6 \nabla^2_2 + 9 \nabla^2_R) Y_L \\
&=& - \Lambda_L Y_L\, , \nonumber
\eea
where
\begin{eqnarray}
\nabla^2_i Y_L &\equiv& \frac{1}{\sin \theta_i} \partial_{\theta_i} (\sin \theta_i \partial_{\theta_i} Y_L) + \Bigl( \frac{1}{\sin \theta_i} \partial_{\phi_i} - \cot \theta_i \partial_\psi \Bigr)^2 \, Y_L\, ,\\
\nabla^2_R Y_L &\equiv& \partial^2_\psi Y_L\, .
\end{eqnarray}
The solution to equation (\ref{equ:eigen}) is obtained through
separation of variables \beq \label{equ:sepvar} Y_L(\Psi) = J_{l_1,
m_1, R}(\theta_1) J_{l_2, m_2, R}(\theta_2) e^{i m_1 \phi_1 + i m_2
\phi_2} e^{\frac{i}{2} R \psi}\, , \eeq where
\begin{equation}
\label{equ:theta}
 \frac{1}{\sin \theta_i} \partial_{\theta_i} (\sin \theta_i \partial_{\theta_i} J_{l_i, m_i, R}(\theta_i)) - \Bigl(\frac{m_i}{\sin \theta_i} - \frac{R}{2} \cot \theta_i \Bigr)^2 J_{l_i, m_i, R}(\theta_i ) =
 - \Lambda_{l_i, R} J_{l_i, m_i, R}(\theta_i)\, .
\end{equation}
The eigenvalues are $\Lambda_{l_i, R} \equiv l_i(l_i+1)-\frac{R^2}{4}$.
Explicit solutions for equation (\ref{equ:theta}) are given in terms of hypergeometric functions ${}_2F_1(a,b,c;x)$
\begin{eqnarray}
J^\Upsilon_{l_i, m_i, R}(\theta_i) &=& N_L^\Upsilon \, (\sin \theta_i)^{m_i} \Bigl( \cot \frac{\theta_i}{2} \Bigr)^{R/2} \times\nonumber\\
&& {}_2F_1 \Bigl(-l_i+m_i,1+l_i+m_i,1+m_i-\frac{R}{2}; \sin^2 \frac{\theta_i}{2} \Bigr) \, ,\\
J^{\Omega}_{l_i, m_i, R}(\theta_i) &=& N_L^{\Omega}\,  (\sin
\theta_i)^{R/2} \Bigl( \cot \frac{\theta_i}{2} \Bigr)^{m_i} \times
\nonumber\\ && {}_2F_1
\Bigl(-l_i+\frac{R}{2},1+l_i+\frac{R}{2},1-m_i+\frac{R}{2}; \sin^2
\frac{\theta_i}{2} \Bigr)\, ,
\end{eqnarray}
where $N_L^\Upsilon$ and $N_L^{\Omega}$ are determined by the
normalization condition (\ref{equ:normal}). If $m_i \ge R/2$,
solution $\Upsilon$ is non-singular. If $m_i \le R/2$, solution
$\Omega$ is non-singular. The full wavefunction corresponds to the
spectrum \beq \Lambda_L = 6 \Bigl( l_1(l_1+1)+l_2(l_2+1) -
\frac{R^2}{8} \Bigr)\, . \eeq The eigenfunctions transform under
$SU(2)_1 \times SU(2)_2$ as the spin $(l_1,l_2)$ representation
and under the $U(1)_R$ with charge $R$. The multi-index $L$ has
the data:
$$ L \equiv (l_1,l_2), (m_1,m_2), R\, .
$$
The following restrictions on the quantum numbers correspond to the existence of
single-valued regular solutions:
\begin{itemize}
\item $l_1$ and $l_2$ are both integers or both half-integers.
\item $ m_1 \in \{-l_1, \cdots, l_1\}$ and $ m_2 \in \{-l_2, \cdots,
l_2\}\, .$
\item $ R \in \mathbb{Z}$  with $\frac{R}{2} \in \{-l_1, \cdots,l_1\}$
and $\frac{R}{2} \in \{-l_2, \cdots, l_2\}$.
\end{itemize}

As discussed in \S\ref{sec:dual}, chiral operators in the dual
gauge theory correspond to $l_1= \frac{R}{2} = l_2$.

\section{Computation of Backreaction in the Singular Conifold}
\label{sec:Calc}

\addtocontents{toc}{\SkipTocEntry}
\subsection{Correction to the Four-Cycle Volume}
Recall the definition (\ref{equ:Delta}) of the (holomorphic)
correction to the warped volume of a four-cycle $\Sigma_4$  \beq
\label{equ:Delta2} \delta V_{\Sigma_4}^w = {\rm Re}(\zeta(X')) =
\int_{\Sigma_4} \d^4 X \sqrt{g^{ind}(X)} \, \delta h(X;X') \, ,
\eeq where $\delta h(X;X') = {\cal C} G(X;X')$ and $T_3 {\cal C} =
2\pi$.

\subsubsection*{Embedding, Induced Metric and a Selection Rule}

The induced
metric on the four-cycle, $g^{ind}$, is determined from the background
metric and the embedding constraint. In \S \ref{sec:ACR} we
introduced the class of supersymmetric embeddings (\ref{equ:Embedding}).

Equation (\ref{equ:psi}) and the form of the angular
eigenfunctions of the Green's function (\S\ref{sec:eigen}) imply
that (\ref{equ:Delta2}) is proportional to \beq \frac{e^{{i \over
2}R \psi_s}}{(2 \pi)^2} \int_0^{2 \pi} \d\phi_1 \, e^{i(m_1 +
\frac{R}{2} n_1) \phi_1} \int_0^{2\pi} \d\phi_2 \, e^{i(m_2 + {R
\over 2} n_2) \phi_2} =  e^{\frac{i}{2}R \psi_s}
\delta_{m_1,-{R\over 2} n_1} \cdot \delta_{m_2,-{R\over 2} n_2}\,
. \eeq We may therefore restrict the computation to values of the
$R$-charge that satisfy \beq \label{equ:mi} m_1 = -{R\over 2}
n_1\, ,\quad m_2 = -{R \over 2} n_2\, . \eeq The winding numbers
$n_i$ (\ref{equ:wind}) are rational numbers of the form \beq
\label{equ:ni} n_i \equiv \frac{\tilde{n}_i}{q}\, , \quad
\tilde{n}_i \in \mathbb{Z}\, , \eeq where $\tilde{n}_i$ and $q$ do
not have a common divisor. Therefore the requirement that the
magnetic quantum numbers $m_i$ be integer or half-integer leads to
the following selection rule for the $R$-charge \beq \label{equ:R}
R = q \cdot k\, , \quad k \in \mathbb{Z}\, . \eeq

\subsubsection*{Green's Function and Reduced Angular Eigenfunctions}

The Green's function on the conifold (\S\ref{subsec:con}) is \beq
G(X;X') = \sum_L Y^*_L(\Psi') Y_L(\Psi) H_L(r;r')\, , \eeq where
it is important that the angular eigenfunctions (\S
\ref{sec:eigen}) are normalized correctly on $T^{1,1}$ \beq
\label{equ:firstnorm} \int \d^5 \Psi \sqrt{g_{T^{1,1}}} |Y_L|^2 =
1\, , \eeq or \beq \label{equ:norm} V_{T^{1,1}} \int_0^1 \d x \,
[J_{l_1,m_1,R}(x)]^2 \int_0^1 \d y \, [J_{l_2,m_2,R}(y)]^2 = 1\, .
\eeq The coordinates $x$ and $y$ are defined in (\ref{equ:xandy}).
Next, we show that the hypergeometric angular eigenfunctions
reduce to Jacobi polynomials if we define \beq l_1 \equiv {R \over
2} + L_1\, , \quad l_2 \equiv {R \over 2} + L_2\, , \quad L_1, L_2
\in \mathbb{Z} \, . \eeq This parameterization is convenient
because chiral terms are easily identified by $L_1 = 0 = L_2$.
Non-chiral terms correspond to non-zero $L_1$ and/or $L_2$.
Without loss of generality we define chiral terms to have $R>0$
and anti-chiral terms to have $R<0$. With these restrictions the
angular eigenfunctions of \S\ref{sec:eigen} simplify to \bea
J_{{R \over 2} + L_1,-{R\over 2}n_1,R}(x) &=&  x^{{R \over 4}(1+n_1)} (1-x)^{{R \over 4}(1-n_1)} \, P_{L_1,R,n_1}(x)\,  ,  \\
J_{{R \over 2} + L_2,-{R\over 2}n_2,R}(y) &=&  y^{{R \over
4}(1+n_2)} (1-y)^{{R \over 4}(1-n_2)} \, P_{L_2,R,n_2}(y) \,  ,
\eea where \bea
P_{L_1,R,n_1}(x) &\equiv& N_{L_1,R,n_1} P_{L_1}^{{R\over 2}(1+n_1),{R\over 2}(1-n_1)}(1-2x)\, ,\\
P_{L_2,R,n_2}(y) &\equiv& N_{L_2,R,n_2} P_{L_2}^{{R\over
2}(1+n_2),{R\over 2}(1-n_2)}(1-2y)\, . \eea The
$P_N^{\alpha,\beta}$ are Jacobi polynomials and the normalization
constants $N_{L_1,R,n_1}$ and $N_{L_2,R,n_2}$ can be determined
from (\ref{equ:norm}).

\subsubsection*{Main Integral}

Assembling the ingredients of the previous subsections (induced
metric, embedding constraint, Green's function) we find that
(\ref{equ:Delta2}) may be expressed as \bea \label{equ:TI}
T_3~\delta V_{\Sigma_4}^w &=& (2\pi)^3 \,
 \int_0^1 \d x \d y\,  \sqrt{ g^{ind}(x,y)}\,  \sum_{L, \psi_s} Y^*_L(x',y') Y_L(x,y) H_L(r;r') \nonumber \\
&=& {V_{T^{1,1}} \over 2} \,  \sum_{L, \psi_s} Y_L^*\,
(r')^{c_L^+} \times e^{{i \over 2}R \psi_s'} r_{\rm
min}^{-c_L^+}\times \frac{I_K^n(Q_L^+)}{\sqrt{\Lambda_L+4}}\, ,
\label{equ:Delta3} \eea where \beq \label{equ:main} I_K^n(Q_L^+)
\equiv \int_0^1 \d x \d y\,  {\cal G}(x,y)\, \left(
\frac{r(x,y)}{r_{\rm min}} \right)^{-6 Q_L^+} P_{L_1,R,n_1}(x)
P_{L_2,R,n_2}(y)\, . \eeq Here $K \equiv (L_1,L_2,R)$, $n \equiv
(n_1,n_2)$ and \beq Q_L^\pm \equiv \frac{c_L^\pm}{6} + {R \over 4}
\, , \quad \quad c_L^\pm \equiv -2 \pm \sqrt{\Lambda_L+4}\, . \eeq
The sum in equation (\ref{equ:Delta3}) is restricted by the
selection rules (\ref{equ:mi}) and (\ref{equ:R}).  Equation
(\ref{equ:main}) is the main result of this section. In the
following we will show that the integral vanishes for all
non-chiral terms and reduces to a simple expression for
(anti)chiral terms.

\addtocontents{toc}{\SkipTocEntry}
\subsection{Non-Chiral Contributions}

In this section we prove that
\bea
\label{equ:magic}
I_K^n(Q) &\equiv& \int_0^1 \d x \d y \ P_{L_1,R,n_1}(x) P_{L_2,R,n_2}(y) \times \nonumber \\
&& \hspace{1cm}  \times x^{Q(1+n_1)} (1-x)^{Q(1-n_1)} y^{Q(1+n_2)} (1-y)^{Q(1-n_2)} \times \nonumber \\
&& \hspace{1cm}\times \Bigl[ \frac{(1+n_1)^2}{2} \frac{1}{x(1-x)} - 2 n_1 \frac{1}{1-x} \nonumber \\
&& \hspace{1.3cm}+ \frac{(1+n_2)^2}{2} \frac{1}{y(1-y)} - 2 n_2
\frac{1}{1-y} -1 \Bigr] \eea vanishes for $Q \to Q_L^+$ iff $L_1
\ne 0$ or $L_2 \ne 0$.
This proves that non-chiral terms do not contribute to the perturbation $\delta V_{\Sigma_4}^w$ to the warped four-cycle volume.\\

The Jacobi polynomial $P_N^{\alpha,\beta}(x)$ satisfies the
following differential equation \bea
&& -N(N+\alpha+\beta+1)P_N^{\alpha,\beta}(1-2x) = \nonumber \\
&& \hspace{1cm} = x^{-\alpha}(1-x)^{-\beta} {d\over dx}\left(x^{1+\alpha}(1-x)^{1+\beta}{d\over dx}P_N^{\alpha,\beta}(1-2x)\right)\, .
\eea
Multiplying both sides by $x^{q_\alpha} (1-x)^{q_\beta}$ and integrating over $x$ gives
\bea
&& -N(N+\alpha+\beta+1)\int_0^1 \d x \, P_N^{\alpha,\beta}(1-2x) x^{q_\alpha}(1-x)^{q_\beta} = \nonumber \\
&& \hspace{1cm} = \int_0^1 \d x \, P_N^{\alpha,\beta}(1-2x) x^{q_\alpha}(1-x)^{q_\beta} \times  \\
&& \hspace{1.5cm} \times \Bigl[
(q_\alpha+q_\beta+1)(\alpha+\beta-q_\alpha-q_\beta)
+{q_\alpha(\alpha-q_\alpha)-q_\beta(\beta-q_\beta)\over
(1-x)}+{q_\alpha(q_\alpha-\alpha)\over x(1-x)}\Bigr] \nonumber\, ,
\eea where we have used integration by parts. In the case of
interest, (\ref{equ:magic}), we make the following
identifications: $N \equiv L_1,\ \alpha \equiv {R\over 2}(1+n_1),\
\beta \equiv {R\over 2}(1-n_1),\ q_\alpha  \equiv Q (1+n_1),\
q_\beta \equiv Q (1-n_1)$. This gives \bea
&& \int_0^1 \d x \, P_{L_1}^{{R\over 2}(1+n_1),{R\over 2}(1-n_1)}(1-2x)\, x^{Q(1+n_1)}(1-x)^{Q(1-n_1)}
\times \left( {(1+n_1)^2\over 2x(1-x)}-{2n_1\over (1-x)}\right) =\nonumber \\
&& = X_{L_1,R,Q}\int_0^1 \d x\, P_{L_1}^{{R\over 2}(1+n_1),{R\over
2}(1-n_1)}(1-2x)\  x^{Q(1+n_1)}(1-x)^{Q(1-n_1)}\, , \eea where
\beq X_{L_1,R,Q} \equiv {(2Q+4Q^2-L_1^2- L_1 R-R-2 L_1 -2RQ)\over
Q(2Q-R)} \nonumber \, . \eeq The corresponding identity for the
$y$-integral follows from the above expression and the
replacements $L_1 \to L_2$ and $n_1 \to n_2$. We then notice that
the integral (\ref{equ:magic}) is \bea
I_K^n(Q) &=& \left(X_{L_1,R,Q}+Y_{L_2,R,Q}-1\right) \times \Lambda_{L_1,R,n_1,Q}\,  \Lambda_{L_2,R,n_2,Q} \nonumber \\
&=& {6(Q-Q_L^+)(Q-Q_L^-)\over Q (2Q-R)} \times \Lambda_{L_1,R,n_1,Q} \, \Lambda_{L_2,R,n_2,Q}\, ,
\eea
where
\bea
\Lambda_{L_1,R,n_1,Q} &\equiv& \int_0^1 \d x\, P_{L_1,R,n_1}(x)\  x^{Q(1+n_1)}(1-x)^{Q(1-n_1)}\, , \label{equ:int1}\\
\Lambda_{L_2,R,n_2,Q} &\equiv& \int_0^1 \d y\, P_{L_2,R,n_2}(y)\
y^{Q(1+n_2)}(1-y)^{Q(1-n_2)}\, . \label{equ:int2} \eea Since
$I_K^n(Q) \propto (Q-Q_L^+)$ it just remain to observe that the
integrals (\ref{equ:int1}) and (\ref{equ:int2}) are finite to
conclude that \beq \lim_{Q \to Q_L^+} I_K^n = 0 \quad {\rm iff}
\quad  Q_L^+ \ne {R \over 2} \, . \eeq This proves that non-chiral
terms do not contribute corrections to the warped volume of any
holomorphic four-cycle of the form (\ref{equ:Embedding}).

\addtocontents{toc}{\SkipTocEntry}
\subsection{Chiral Contributions}

Finally, let us consider the special case $Q_L^+ = {R \over 2}$
which corresponds to chiral operators ($L_1=L_2=0$) in the dual
gauge theory. In this case, \beq I_R^{\rm chiral} \equiv \lim_{Q
\to {R \over 2}} I_K^n = \frac{3R+4}{2} \frac{1}{R}  \times
\Lambda_{0,R,n_1,{R \over 2}}\times  \Lambda_{0,R,n_2,{R \over
2}}\, , \eeq where \bea
\Lambda_{0,R,n_1,{R \over 2}} &\equiv& \int_0^1 \d x \, P_{0,R,n_1}(x) \  x^{{R \over 2}(1+n_1)}(1-x)^{{R \over 2}(1-n_1)}\, ,\\
\Lambda_{0,R,n_2,{R \over 2}} &\equiv& \int_0^1 \d y \,
P_{0,R,n_2}(y) \  y^{{R \over 2}(1+n_2)}(1-y)^{{R \over
2}(1-n_2)}\, . \eea Notice that $P_{0,R,n_i} = N_{0,R,n_i} =
(N_{0,R,n_i})^{-1}(P_{0,R,n_i})^2$. Hence, \bea
\Lambda_{0,R,n_1,{R \over 2}} &\equiv& (N_{0,R,n_1})^{-1}\int_0^1 \d x \left( P_{0,R,n_1}(x)  \left[x^{(1+n_1)}(1-x)^{(1-n_1)}\right]^{R/4} \right)^2 \nonumber \\
\Lambda_{0,R,n_2,{R \over 2}} &\equiv& (N_{0,R,n_2})^{-1}\int_0^1
\d y \left( P_{0,R,n_2}(y)
\left[y^{(1+n_2)}(1-y)^{(1-n_2)}\right]^{R/4} \right)^2 \nonumber
\eea and \beq \Lambda_{0,R,n_1,{R \over 2}} \times
\Lambda_{0,R,n_2,{R \over 2}} = \frac{1}{V_{T^{1,1}} N_{0,R,n_1}
N_{0,R,n_2}} \eeq by the normalization condition (\ref{equ:norm})
on the angular wave function. Therefore, we get the simple result
\beq \frac{I_R^{\rm chiral}}{\sqrt{\Lambda_R^{\rm chiral}+4}} =
\frac{1}{V_{T^{1,1}} N_{0,R,n_1} N_{0,R,n_2}} \times \frac{1}{R}
\, . \eeq We substitute this into equation (\ref{equ:Delta3}) and
get \beq \label{equ:delta} T_3~ (\delta V_{\Sigma_4}^w)_{\rm
chiral} =  {1 \over 2} \sum_s \sum_{R = q \cdot k} \frac{1}{R}
\times  \Bigl(\prod_i (\bar{w}_i')^{p_i}\Bigr)^{R /P} \times {1
\over \bar{\mu}^{R}} \times e^{i {R \over P} 2\pi s}  \, , \eeq
where we used \beq (r')^{3R/2} \frac{Y_R^*(\Psi')}{N_{0,R,n_1}
N_{0,R,n_2}} = \Bigl(\prod_i (\bar{w}_i')^{p_i}\Bigr)^{R/P} \eeq
and \beq e^{i {\rm arg}(\mu)R} r_{\rm min}^{-3R/2} = {1 \over
\bar{\mu}^R}\, . \eeq The sum over $s$ in (\ref{equ:delta}) counts
the $P$ different roots of equation (\ref{equ:Embedding}): \beq
\sum_{s=0}^{P-1} e^{{q \cdot k \over P} 2 \pi s} = P \, \delta_{{q
\cdot k \over P},j}\, , \quad j \in \mathbb{Z}\, . \eeq Dropping
primes, we therefore arrive at the following sum \beq T_3~ (\delta
V_{\Sigma_4}^w)_{\rm chiral} = {1 \over 2} \sum_{j=1}^\infty
\frac{1}{j} \times \Bigl(\prod_i \bar{w}_i^{p_i}\Bigr)^{j} \times
{1 \over \bar{\mu}^{P \cdot j}}\, , \eeq which gives \beq T_3~
(\delta V_{\Sigma_4}^w)_{\rm chiral} = - \frac{1}{2} \, \log
\left[1-\frac{\prod_i \bar{w}_i^{p_i}}{\bar{\mu}^P}\right]\, .
\eeq

For the anti-chiral terms ($R < 0$) an equivalent computation
gives the complex conjugate of this result.

The $R=0$ term formally gives a divergent contribution that needs
to be regularized by introducing a UV cutoff at the end of the
throat. Alternatively, as discussed in \S\ref{sec:dual}, this term
does not appear if we define $\delta h$ as the solution of
(\ref{equ:laplace}) with $\sqrt{g}\, \rho_{bg}(Y) =
\delta^{(6)}(Y-X_0)$. This choice amounts to evaluating the change
in the warp factor, $\delta h$, created by moving the D3-brane
from some reference point $X_0$ to $X$. We may choose the
reference point $X_0$ to be at the tip of the cone, $r=0$, and
thereby remove the divergent
zero mode.\\

The total change in the warped volume of the four-cycle is
therefore \beq \delta V_{\Sigma_4}^w= (\delta V_{\Sigma_4}^w)_{\rm
chiral} + (\delta V_{\Sigma_4}^w)_{\rm anti-chiral} \eeq and \beq
T_3 \, {\rm Re} (\zeta) = T_3\, \delta V_{\Sigma_4}^w= - {\rm Re}
\Bigl(  \log \left[\frac{ \mu^P - \prod_i w_i^{p_i} }{\mu^P}
\right] \Bigr) \, . \eeq Finally, the prefactor of the
nonperturbative superpotential is \beq A(w_i) =A_0\, e^{-T_3 \zeta
/n } = A_0\,  \Bigl(\frac{\mu^P - \prod_i w_i^{p_i}
}{\mu^P}\Bigr)^{1/n}\, . \eeq

\section{Computation of Backreaction in $Y^{p,q}$ Cones}
\label{sec:ypq}

\addtocontents{toc}{\SkipTocEntry}
\subsection{Setup}

\subsubsection*{Metric and Coordinates on $Y^{p,q}$}

Cones over $Y^{p,q}$ manifolds have the metric \beq \d s^2= \d
r^2+r^2 \d s^2_{Y^{p,q}} \,, \eeq where the Sasaki-Einstein metric
on the $Y^{p,q}$ base is given by \cite{Gauntlett2,Gauntlett} \bea
\d s^2_{Y^{p,q}} &=& {1-y\over6}\,(\d\theta^2+\sin^2\theta
\,\d\phi^2) + {1\over
v(y)w(y)} \d y^2 + {v(y)\over 9} (\d\psi + \cos\theta \, \d\phi)^2 \nonumber \\
&& + w(y) \big[\d\alpha + f(y) \left( \d\psi + \cos\theta \,
\d\phi \right) \big]^2\, . \label{alphametric} \eea The following
functions have been defined: \beq v(y) \equiv {b - 3y^2 + 2y^3
\over b - y^2}\,, \quad w(y) \equiv {2 (b - y^2) \over 1 - y}\,,
\quad f(y) \equiv {b - 2y + y^2 \over 6 (b - y^2)}\, , \eeq with
\beq b = {1\over 2}-{p^2-3q^2\over 4p^3}\,\,\sqrt{4p^2-3q^2}\,.
\eeq The parameters $p$ and $q$ are two coprime positive integers.
The zeros of $v(y)$ are \beq y_{1,2}\equiv {1\over 4p}\,
\Big(\,2p\, \mp\,3q-\sqrt{4p^2-3q^2}\,\Big)\, , \qquad y_3 \equiv
{3\over 2}-(y_1+y_2)\, . \eeq It is also convenient to introduce
\beq x ={y-y_1\over y_2-y_1}\ . \eeq The angular coordinates
$\theta$, $\phi$, $\psi$, $x$, and $\alpha$ span the ranges: \bea
&& 0 \le \theta\le \pi\, , \quad 0< \phi\le
2\pi\, ,\quad 0< \psi\le 2\pi\, , \nonumber \\
&& 0\le x\le 1\, ,\quad  0< \alpha\le 2\pi \ell\, , \eea where
$\ell \equiv -{q\over 4p^2y_1y_2}$.

\subsubsection*{Green's Function}

The Green's function on the $Y^{p,q}$ cone is
\beq
G(X;X') = \sum_L \frac{1}{4(\lambda+1)}\times
Y_L^*(\Psi') Y_L(\Psi) \times
\begin{cases}
\frac{1}{r'^4} \Bigl(\frac{r}{r'}\Bigr)^{2\lambda} & r \le r'\, , \\
\frac{1}{r^4} \Bigl(\frac{r'}{r}\Bigr)^{2\lambda} & r \ge r'\, .
\end{cases}
\eeq Here $L$ is again a complete set of quantum numbers and
$\Psi$ represents the set of angular coordinates
$(\theta,\phi,\psi,x,\alpha)$. The eigenvalue of the angular
Laplacian is $\Lambda_L \equiv 4\lambda(\lambda+2)$. The spectrum
of the scalar Laplacian on $Y^{p,q}$, as well as the
eigenfunctions $Y_L(\Psi)$, were calculated in
\cite{Berenstein,Kihara:2005nt}.  We do not review this treatment
here, but simply present an explicit form of $Y_L(\Psi)$ \beq
Y_L(\Psi)= N_L \, e^{i(m\phi+n_\psi \psi+{n_\alpha\over
\ell}\alpha)}J_{l,m,2n_\psi}({\theta})R_{n_\alpha,n_\psi,l,\lambda}(x)
\, ,\label{equ:YpqA} \eeq where \bea
R_{n_\alpha,n_\psi,l,\lambda}(x)=x^{\alpha_1}(1-x)^{\alpha_2}(a-x)^{\alpha_3}h(x)\,
, \quad a \equiv {y_1-y_3\over y_1-y_2}\ . \eea The parameters
$\alpha_i$ depend on $n_\psi,n_\alpha$ (see \cite{Kihara:2005nt}),
and the function $h(x)$ satisfies the following differential
equation \beq \label{equ:Heunsequation} \left[{d^2\over
dx^2}+\left({\gamma\over x}+{\delta \over x-1}+{\epsilon\over x-a
}\right){d\over dx}+{\alpha \beta x-k\over
x(1-x)(a-x)}\right]h(x)=0\, . \eeq The parameters $\alpha$,
$\beta$, $\gamma$, $\delta$, $\epsilon$, $k$ depend on $p,q$ and
on the quantum numbers of the $Y^{p,q}$ base. Explicit expressions
may be found in \cite{Kihara:2005nt}.

Finally, we introduce the normalization condition that fixes $N_L$
in (\ref{equ:YpqA}). If we define $z \equiv \sin^2{\theta\over 2}$
then the normalization condition \beq \int \d^5
\Psi\sqrt{g_{Y^{p,q}}} |Y_L|^2 = 1 \eeq becomes \beq
\label{equ:normalization} N_L^2 \int_0^1 \d z \,\d x\,
\sqrt{g(x,z)} \, J^2 R^2 = {1\over (2\pi)^3\ell}\, , \eeq where
\beq \sqrt{g(x,z)}=\sqrt{g(x)}={q(2p+3q+\sqrt{4p^2-3q^2}-6qx)
\over 24p^2}\ . \eeq

\subsubsection*{Embedding, Induced Metric and a Selection Rule}

The holomorphic embedding of four-cycles in $Y^{p,q}$ cones is described by the
algebraic equation \cite{ypqsusybranes} \beq \prod_{i=1}^3
w_i^{p_i} = \mu^{2p_3}\, , \eeq where \bea
w_1 &\equiv& \tan{\theta\over 2}\, e^{-i\phi}\,,\\
w_2 &\equiv& \frac{1}{2}{\sin \theta}\, x^{{1\over 2y_1}}(1-x)^{{1\over 2y_2}}(a-x)^{{1\over 2y_3}}e^{i(\psi+6\alpha)}\,,\\
w_3 &\equiv& \frac{1}{2}r^3\, {\sin\theta}\,
[x(1-x)(a-x)]^{1/2}e^{i\psi}\, . \eea This results in the
following embedding equations in terms of the real coordinates
\bea \label{equ:Embedding1} \psi &=& {1 \over
1+n_2} \left(n_1 \phi - 6 n_2 \alpha \right) - \psi_s \,, \\
r &=&r_{\rm min}\left[z^{1+n_1+n_2}(1-z)^{1-n_1+n_2} \right]^{-1/6} \left[x^{2e_1}(1-x)^{2e_2}(a-x)^{2e_3}\right]^{-1/6} \nonumber \\
&\equiv& r_{\rm min} r_z r_x\, , \eea where \bea \psi_s &\equiv&
{\rm arg}(\mu)+{2\pi s \over p_2+p_3}\,, \quad
s\in\{0,1,\dots,(p_2+p_3)-1\}
\\  r^{3/2}_{\rm min} &\equiv&|\mu|\,,
\eea and \bea
e_i &\equiv& {1\over 2}\left(1+{n_2\over y_i}\right)\, ,\\
n_1 &\equiv&{p_1\over p_3}\,, \\
n_2 &\equiv& {p_2\over p_3}\ . \eea

Integration over $\phi$ and $\alpha$ together with the embedding
equation (\ref{equ:Embedding1}) dictates the following selection
rules for the quantum numbers of the angular eigenfunctions
(\ref{equ:YpqA}), \bea m=-{n_1\over 2}Q_R\ ,\ \ \ n_\alpha=3\ell n_2 Q_R %
\ ,\ \ \  n_\psi={1+n_2\over 2}Q_R\, , \eea where $Q_R$ is the
$R$-charge defined as $Q_R \equiv 2n_\psi-{1\over 3\ell}n_\alpha$.
In this case $\alpha_i=e_i{Q_R\over 2}$.

Finally, we need the determinant of the induced metric on the
four-cycle \beq \d\theta \d x\, \sqrt{g^{ind}}={r^4\over
z(1-z)x(1-x)(a-x)}\, \mathcal{G}(x,z) \, \d z \d x\, . \eeq The
function $\mathcal{G}$ is too involved to be written out
explicitly here, but is available upon request.  It is a
polynomial of order $3$ in $x$ and of order $2$ in $z$.

\subsubsection*{Main Integral}

The main integral  (the analog of (\ref{equ:main})) is therefore
given by \beq I_L=\int {\d x \d z \, \mathcal{G}(x,z) \, N_L^2
\over z(1-z)x(1-x)(a-x)}\left({r\over r_{\rm
min}}\right)^{-6Q_L^+}P^{a,b}_{A=l-n_\psi}(1-2z)h_L(x) \, ,\eeq
with $a \equiv (1+n_1+n_2){Q_R\over 2}$, $b
\equiv(1-n_1+n_2){Q_R\over 2}$ and $6Q_L^+ \equiv 2\lambda+{3\over
2}Q_R$. We will calculate this integral for a general
$6Q_L^+=2w+{3\over 2}Q_R$ and then take the limit $w \to \lambda$.

First we compute the integral over $z$ in complete analogy to the
treatment of Appendix \ref{sec:Calc}. The Jacobi polynomial
satisfies \beq \label{equ:Jbi_equ2}
 r_z^{3Q_R}{d\over
dz}\Bigl(r_z^{-3Q_R}z(1-z){d\over dz}P^{a,b}_A(1-2z)\Bigr)+
A(A+1+a+b)P^{a,b}_A(1-2z)=0\, . \eeq Let us multiply this equation
by $r_z^{-(2w+{3\over 2}Q_R)}$ and integrate over $z$. It can be
shown that there is a third order polynomial $\mathbb{G}(x)$ which
is implicitly defined by the following relation \bea
\label{equ:int_z2} && {\mathcal{G}(x,z)\over
z(1-z)}-\mathbb{G}(x)={\mathcal{G}(x,z=0)\over(1+n_1+n_2)^2\left({w^2\over
9^2}-{Q_R^2\over 16}\right)} \times \nonumber \\
&& \times \left[ r_z^{2w+{3\over 2}Q_r}{d\over
dz}\left(z(1-z)r_z^{-3Q_R}{d\over dz}\left(r_z^{{3\over
2}Q_R-2w}\right)\right)+A(A+1+a+b) \right]\, . \eea The right-hand
side vanishes after multiplying by $r_z^{-6Q_L^+}P^{a,b}_A(1-2z)$
and integrating, and we get \beq I_L=\int{\d x\, \mathbb{G}(x)\,
N_L^2 \over x(1-x)(a-x)}\, r_x^{-6Q_L^+}h_L(x)\int \d z \,
r_z^{-6Q_L^+} P_A^{a,b}(1-2z) \ . \label{equ:Int3} \eeq

\addtocontents{toc}{\SkipTocEntry}
\subsection{Non-Chiral Contributions}

To evaluate (\ref{equ:Int3}) we make use of the differential
equation (\ref{equ:Heunsequation}). We multiply
(\ref{equ:Heunsequation}) by $r_x^{-2w-{3\over 2}Q_R}$ and
integrate over $x$. There exists a first order polynomial
$M\sqrt{g(x)}$ such that \bea \label{equ:imp} &&
{\mathbb{G}(x)\over x(1-x)(a-x)}-M\sqrt{g(x)}
=\nonumber\\
&=& {144\, \mathbb{G}(x=0)\over (1-n_2)(3Q_R+4\lambda)(18Q_R
n_2+8\lambda n_2-9Q_R-4\lambda-24)} \times \Bigl[(\alpha\beta
x-k)-  \nonumber \\
&& -r_x^{2w+{3\over 2}Q_R}{d\over dx}\Bigl( r_x^{-2w-{3\over
2}Q_R}(\gamma (1-x)(a-x)+\delta x(x-a)+\epsilon x(x-1))\Bigr)
 \nonumber \\
&& +r_x^{2w+{3\over 2}Q_R}{d^2\over
dx^2}\Bigl(x(1-x)(a-x)r_x^{-2w-{3\over 2}Q_R}\Bigr)\Bigr]\, , \eea
where we defined \beq \label{equ:M}
 M \equiv {48(\lambda-w)(\lambda+w+2)\over(1+n_2)(16
w^2-9Q_R^2 )}\ . \eeq After multiplying by $r_x^{-6Q_L^+}h(x)$ and
integrating over $x$, the right-hand side vanishes and we have
\bea I_L &=& M N_L^2\int \d x \d z\, \sqrt{g(x,z)} \left({r\over
r_{\rm min}}\right)^{-6Q_L^+} P_A^{a,b}(1-2z)h(x)
\\ &=& M N_L\int \d z \d x\, \sqrt{g} \, \left({r\over r_{\rm min}}\right)^{-2\lambda}\,
J R \, . \eea Since $\lim_{w \to \lambda} M = 0$, this immediately
implies that $\lim_{w \to \lambda}I_L=0$ `on-shell', {\it{i.e.}}
for all operators except for the chiral ones. Just as for the
singular conifold case, we have therefore proven that non-chiral
terms do not contribute to the perturbation to the warped
four-cycle volume.

\addtocontents{toc}{\SkipTocEntry}
\subsection{Chiral Contributions}

For the chiral operators one finds \beq \lambda={3\over 4}Q_R\ \eeq and both the numerator and the
denominator of $M$ (\ref{equ:M}) vanish. Chiral operators also
require $A=l-n_{\psi}$ to be equal to zero. Taking the chiral limit we therefore
find
 \bea \label{equ:IL} I_L&=&{(3Q_R+4)\over (1+n_2)Q_R}N_L^2 \int
{\d x\, q(2p+3q+\sqrt{4p^2-3q^2}-6qx) \over 24p^2}\left({r\over
r_{\rm \min}}\right)^{-{3}Q_R}  \\&=&{(3Q_R+4)\over
(1+n_2)Q_R}{1\over (2\pi)^3 \ell}\, , \eea since $A=0$ implies
$P_A^{a,b}(1-2z)=1$ and $h(x)=1$. The integral in (\ref{equ:IL})
reduces to the normalization condition (\ref{equ:normalization}).
Finally, we use the identity for chiral wave-functions $r^{{3\over
2}Q_R}Y_L(\Psi)=\left(w_1^{n_1} w_2^{n_2} w_3\right)^{{Q_R\over
2}}$ and the relation between $T_3 (\delta V_{\Sigma_4}^w)_{\rm
chiral}$ and $I_L$ (an analog of (\ref{equ:TI})). Note that the
$(2\pi)^3$ in (\ref{equ:TI}) should be changed to $(2\pi)^3 \ell$
as $\alpha$ runs from $0$ to $2\pi \ell$.  We hence arrive at the
analog of (\ref{equ:delta}) \bea T_3 (\delta V_{\Sigma_4}^w)_{\rm
chiral} ={1\over 2}\sum_{Q_R, s} {2\over
(1+n_2)Q_R}\left(\bar{w}_1^{n_1} \bar{w}_2^{n_2}
\bar{w}_3\right)^{{Q_R\over 2}}e^{i{(1+n_2)\over 2}Q_R \psi_s}\, ,
\eea where we recall that $\psi_s={2\pi s\over p_2+p_3}$. The
summation over $s$ effectively picks out $n_\psi={(1+n_2)\over
2}Q_R$ to be of the form $(p_2+p_3) s'$ with natural
 $s'$, or
$Q_R=2p_3 s'$. After summation over $s'$ we have \beq
\label{equ:chiralypqans} T_3 (\delta V_{\Sigma_4}^w)_{\rm chiral}
= -{1\over 2} \log\left[{\bar{\mu}^{2p_3}-\prod_i
\bar{w}_i^{p_i}\over \bar{\mu}^{2p_3}} \right]\, . \eeq

A similar calculation for the anti-chiral contributions gives the
complex conjugate of (\ref{equ:chiralypqans}).\\

The final result for the perturbation of the warped volume of
four-cycles in cones over $Y^{p,q}$ manifolds is then \beq T_3\,
\delta V_{\Sigma_4}^w= - {\rm Re} \Bigl(  \log \left[\frac{
\mu^{2p_3} - \prod_i w_i^{p_i} }{\mu^{2p_3}} \right] \Bigr) \, ,
\eeq so that \beq A(w_i) = A_0\, \Bigl(\frac{\mu^{2 p_3} - \prod_i
w_i^{p_i} }{\mu^{2 p_3}}\Bigr)^{1/n}\, . \eeq

\newpage

\begingroup\raggedright\endgroup

\end{document}